\documentclass[12pt]{article}
\usepackage{amssymb,amsmath,epsfig,graphics}
\allowdisplaybreaks

\begin{document}

\title{\bf Traversable Wormhole Solutions admitting Noether Symmetry in $f(\mathcal{R,}\mathcal{T}^{2})$ Theory}
\author{M. Zeeshan Gul \thanks{mzeeshangul.math@gmail.com} and M. Sharif \thanks {msharif.math@pu.edu.pk}\\
Department of Mathematics and Statistics, The University of Lahore,\\
1-KM Defence Road Lahore, Pakistan.}

\date{}
\maketitle

\begin{abstract}
This paper uses the Noether symmetry approach to examine the viable
and stable traversable wormhole solutions in the framework of
$f(\mathcal{R,}\mathcal{T}^{2})$ theory, where $\mathcal{R}$ is the
Ricci scalar and
$\mathcal{T}^{2}=\mathcal{T}_{\mu\nu}\mathcal{T}^{\mu\nu}$ is the
self-contraction of stress-energy tensor. For this purpose, we
consider a specific model of this modified theory to obtain exact
solutions of the Noether equations. Further, we formulate the
generators of Noether symmetry and first integrals of motion. We
analyze the presence of viable and stable traversable wormhole
solutions corresponding to different redshift functions. In order to
determine whether this theory provides physically viable and stable
wormhole geometry or not, we check the graphical behavior of null
energy constraint, causality condition and adiabatic index for
effective stress-energy tensor. It is found that viable and stable
traversable wormhole solutions exist in this modified theory.
\end{abstract}

\section{Introduction}

The general theory of relativity (GR) is the most effective theory
of gravity which describes a wide range of gravitational effects
from small to large structures in the cosmos. This theory passes the
solar system tests successfully. Recent observations confirmed the
existence of gravitational waves and showed that their power
spectrum and attributes are consistent with those predicted by GR.
The most comprehensive model for explaining the dynamics of the
cosmos is the CDM model. But the cosmological constant in the action
of GR leads to the cosmological constant problem. However, there are
many other unresolved issues such as the dark energy paradox and the
existence of singularities which keep open the way to extend GR. It
is fascinating that modifying GR can help in finding solutions of
all these issues. There are various modified gravitational theories
that successfully describe the mysterious universe. The
$f(\mathcal{R})$ gravity is simplest modified theory whose useful
literature has been made available to comprehend the realistic
aspects of this theory \cite{1}.

The prediction of singularities at high energy level where GR is not
applicable because of possible quantum effects is considered a
significant issue in GR. But quantum gravity does not have a
specific formalism. Accordingly, a new gravitational theory has been
established by adding non-linear term
$\mathcal{T}^{\mu\nu}\mathcal{T}_{\mu\nu}$ in the integral action
named as $f(\mathcal{R},\mathcal{T}^{2})$ gravity \cite{2}. This
theory is also equivalent to GR in a vacuum. This modified theory is
assumed as the most successful approach to resolve the spacetime
singularity in the non-quantum description. This theory is also
known as energy-momentum squared gravity (EMSG). Thus, the field
equations of EMSG deviate from GR in the presence of matter source.
This theory includes higher-order matter and curvature terms in the
field equations which are used to examine several interesting cosmic
consequences. It is noteworthy that this modified proposal resolves
the spacetime singularity but cosmic evolution remains unaffected.

This mathematical model does not support big-bang theory as the
scale factor is minimum and the maximum energy density is finite in
the early times. But the density profile in the radiation-dominated
era manifests that EMSG favors the inflationary cosmic models. These
models resolves main cosmic problems such as flatness and horizon
issues but no model of inflation has been confirmed by observations.
A class of cosmic models (varying the speed of light theories) have
been developed in this perspective which does not support the
inflation. This suggested an alternative way to solve these cosmic
problems by varying the speed of light and Newton's constant of
gravitation. Theories about varying the speed of light were
motivated to resolve the inflation problems but do not resolves the
big bang singularity. To address this problem, Bhattacharjee and
Sahoo \cite{2a} proposed a novel cosmic model which is free from
inflation as well as big-bang singularity by including a Newton's
constant of gravitation and varying the speed of light in the
context of EMSG. Singh et al \cite{2b} studied the viability and
stability of color-flavor locked quark stars in this framework.
Nazari \cite{2c} examined that this theory passes the solar system
tests successfully and found that except for a small deviation, the
behavior of light curves in EMSG is similar to GR.

The presence of $\mathcal{T}^{2}$ term yields some quadratic
corrections to the Friedman equations which are similar to those
reported in the framework of loop quantum gravity \cite{3a}. Board
and Barrow \cite{3b} analyzed a range of exact solutions for
isotropic universe and examined their behavior through accelerated
expansion and the presence or absence of singularities. Akarsu et al
\cite{3c} proposed energy-momentum powered gravity by adding a term
$f(\mathcal{T}_{\mu\nu}\mathcal{T}^{\mu\nu})$ in functional action
and discussed a specific case
$f(\mathcal{T}_{\mu\nu}\mathcal{T}^{\mu\nu})=
\alpha(\mathcal{T}_{\mu\nu}\mathcal{T}^{\mu\nu})^{\eta}$, where
$\alpha$ and $\eta$ are real constants. They analyzed that this
theory can be unified with Starobinsky gravity to explain the
complete cosmic history including inflationary era. Akarsu et al
\cite{3d} established a scale-independent EMSG that lead to
scenarios with many interesting applications in cosmology. Ranjit et
al \cite{3e} investigated solutions for matter density and studied
their cosmic consequences in EMSG. Sharif and Naz \cite{3f}
investigated viable features of a gravastar in this framework.

Chen and Chen \cite{4a} investigated the axial perturbations of the
charged black holes in EMSG theory. It is worthwhile to mention here
that this theory is not limited to bouncing solutions and the early
universe. But, this can be used to manipulate the CMB temperature
fluctuation \cite{4b}. Kazemi et al \cite{4c} analyzed the
gravitational stability of an infinite fluid as well as
differentially rotating fluid in this framework. Rudra and
Pourhassan \cite{4e} explored the thermodynamic properties of a
black hole in the EMSG. Nazari et al \cite{4f} examined the Palatini
formulation of EMSG and studied their consequences in various
contexts. We have studied the stability of the Einstein universe
\cite{5}-\cite{5a} dynamics of relativistic objects
\cite{6}-\cite{6d} in this framework. Yousaf et al \cite{4g}
analyzed the effects of EMSG on the dynamics of axially symmetric
anisotropic and dissipative fluid. Khodadi and Firouzjaee \cite{4h}
used the linear perturbations on Reissner-Nordstrom-de Sitter
solutions in this framework and developed the valid study of cosmic
censorship conjecture beyond Einstein's gravity.

The surprising and ambiguous characteristics of our cosmos put
forward stunning questions for the scientific community. The
existence of hypothetical structures is considered the most
significant issue that gives the wormhole (WH) structure. It is
described as a hypothetical bridge that connects two distinct parts
of the universe due to the presence of exotic matter (which violates
energy conditions). The intra-universe WH connects different regions
of the same cosmos while inter-universe WH joins two distinct parts
of different cosmos. Flamm \cite{10} developed WH structure through
the Schwarzschild solution. Later, Einstein and Rosen \cite{11}
found that a curved space structure can join two different
spacetimes through a bridge named as Einstein-Rosen bridge. Wheeler
\cite{12} examined that Schwarzschild WH is non-traversable because
two-way traveling is not possible in it, and anything attempting to
pass through would be destroyed by the tremendous tidal forces
present at the WH throat. Moreover, the WH throat rapidly expands
from zero to a finite circumference and compresses to zero with
time, and prevents the access to anything. However, Fuller and
Wheeler \cite{13} investigated that WHs would collapse instantly
after the formation.

The maximum amount of exotic matter in the bridge raises questions
about the existence of a viable WH structure. Thus, there should not
be an excessive amount of exotic matter in the bridge for viable WH
geometry. The first traversable WH was proposed by Morris and Thorne
\cite{14}. In addition to the existence of such hypothetical
structures, stability is the most significant issue that describes
how these cosmic structures respond to perturbations and enhances
their physical features. However, a stable state is obtained due to
non-singular configuration which prevents the WH from collapsing in
contrast to unstable WH, that can also exist because of very slow
decay. Several methods have been established to investigate the
viable and stable WH structures \cite{15}. Dzhunushaliev et al
\cite{16} investigated the stability of WH configuration with and
without electromagnetic field. Oliveira et al \cite{16a} examined
physically viable and stable traversable Yukawa-Casimir WHs.

Symmetry describes the properties of mathematical and physical
systems that remain invariant due to perturbation. The uses of
symmetry techniques are significant for obtaining viable solutions
to differential systems. The continuous symmetry (which occurs due
to constant change in a system) corresponding to the Lagrangian is
known as Noether symmetry (NS). The associated Lagrangian is useful
to identify the realistic aspects of a physical system by providing
information about various symmetries of the system. However, NS
technique is the most elegant approach that describes a connection
between NS generators and conserved quantities of the system
\cite{17}. The complexity of the system is reduced by this method
and viable solutions are obtained that can be used to study the dark
cosmos. The literature provides several ways to explain the NS
methodology \cite{18}. For example, one way to identify the symmetry
generators is Noether gauge symmetry in which gauge term is added to
the invariance condition, while another method is to set the Lie
derivative of the Lagrangian to zero. This technique also produces
some useful restrictions that allow one to select cosmological
models according to the recent observations \cite{19}.

Noether charges play a significant role as they are used to
investigate several cosmic issues in various backgrounds. Motavali
and Golshani \cite{20} used the NS method to obtain exact
cosmological solutions of FRW spacetime. Vakili \cite{21} used this
approach to analyze dark components of the universe. Capozziello et
al \cite{22} analyzed this strategy in the quintessence and phantom
cosmic models. Capozziello et al \cite{23} obtained viable solutions
of static spherically spacetime through the NS method in
$f(\mathcal{R})$ theory. Shamir et al \cite{24} obtained the exact
solutions of the FRW universe model in the same theory. Jamil et al
\cite{25} investigated scalar field cosmology through the NS
approach in teleparallel theory. Momeni et al \cite{26} studied
exact cosmological solutions through NS in
$f(\mathcal{R},\mathcal{T})$ theory. Shamir and Ahmad \cite{27}
examined isotropic and anisotropic solutions via NS technique in
$f(\mathcal{G},\mathcal{T})$ theory. We have found exact solutions
through the NS technique in $f(\mathcal{R},\mathcal{T}^2)$ theory
\cite{28}.

Cosmologists have been quite passionate about studying WH geometry
in modified theories. Lobo et al \cite{29} examined traversable WH
structure through distinct types of WH shape functions (WSFs) and
equations of state in $f(\mathcal{R})$ theory. Mazharimousavi and
Halilsoy \cite{31} found that WH solutions satisfy all the necessary
requirements near the WH throat in this theory. In the framework of
scalar-tensor theory, the traversable WH geometry through NS has
been examined in \cite{32}. The static WH solutions with different
matter contents in $f(\mathcal{R},\mathcal{T})$ theory has been
analyzed in \cite{33}. The viable WH solutions admitting NS in
$f(\mathcal{R})$ theory has been studied in \cite{34}. Sharif et al
\cite{34a} studied new holographic dark energy model and Tsallis
holographic dark energy model in the context of modified theories of
gravity. Mustafa et al \cite{35} analyzed viable WH geometry through
the Karmarkar condition in $f(Q)$ theory. Shamir and Fayyaz
\cite{36} developed a WSF through embedding class-I technique in
$f(\mathcal{R})$ theory and examined that WH structure can be
obtained with a negligible amount of exotic matter. Hassan et al
\cite{36a} found that WH solutions corresponding to linear and
exponential model of $f(Q)$ gravity models are physically viable and
stable. Malik et al \cite{37} used the Karmarkar condition to study
traversable WH structure in $f(\mathcal{R})$ theory.

The above literature motivates us to examine WH geometry through NS
approach in $f(\mathcal{R},\mathcal{T}^2)$ theory. The paper is
designed in the following way. In Section \textbf{2}, we develop the
field equations of static spherical spacetime in
$f(\mathcal{R},\mathcal{T}^2)$ theory. Section \textbf{3} gives a
brief discussion of WH solutions via the NS technique. In Section
\textbf{4}, we analyze the stability of WH solutions by
\emph{causality condition} and \emph{adiabatic index}. The last
section summarize our results.

\section{Basic Formalism of $f(\mathcal{R},\mathcal{T}^2)$ Theory}

This modified theory is defined by the following action \cite{2}
\begin{equation}\label{1}
\mathcal{A}=\int(\frac{f(\mathcal{R}, \mathcal{T}^{2})}{2\kappa}+
L_{m})\sqrt{-g}d^4x,
\end{equation}
where $L_{m}$, $\kappa$ and $g$ represent the matter-Lagrangian,
coupling constant and determinant of the line element, respectively.
The corresponding field equations are
\begin{equation}\label{2}
(g_{\mu\nu}\nabla_{\mu}\nabla^{\mu}+\mathcal{R}_{\mu\nu}-\nabla
_{\mu}\nabla_{\mu})f_{\mathcal{R}}-\frac{1}{2}g_{\mu\nu }f =
\mathcal{T}_{\mu\nu}-\Theta_{\mu\nu}f_{ \mathcal{T}^{2}},
\end{equation}
where $f\equiv f(\mathcal{R}, \mathcal{T}^{2})$,
$f_{\mathcal{T}^{2}}= \frac{\partial f} {\partial \mathcal{T}^{2}}$,
$f_{\mathcal{R}}= \frac{\partial f} {\partial \mathcal{R}}$, and
\begin{eqnarray}\label{3}
\Theta_{\mu\nu}
=2\mathcal{T}_{\mu}^{\xi}\mathcal{T}_{\nu\xi}-\mathcal{T}\mathcal{T}
_{\mu\nu}-2L_{m}\mathcal{T}_{\mu\nu}+L_{m}g_{\mu\nu}\mathcal{T}
-\frac{4\partial^{2}L_{m}}{\partial g^{\mu\nu}\partial
g^{\xi\eta}}\mathcal{T}^{\xi\eta}.
\end{eqnarray}
We assume isotropic fluid configuration as
\begin{equation}\label{4}
\mathcal{T}_{\mu\nu}=\mathcal{U}_{\mu}\mathcal{U}_{\nu}\rho
+p(\mathcal{U}_{\mu}\mathcal{U}_{\nu}+g_{\mu\nu}).
\end{equation}
Using this value in Eq.(\ref{3}), we have
\begin{eqnarray}\label{4a}
\Theta_{\mu\nu} =-(4p\rho+\rho^2+3p^2)\mathcal{U}_{\mu}\mathcal{U}
_{\nu}.
\end{eqnarray}
Re-arranging Eq.(\ref{2}), we obtain
\begin{equation}\label{5}
G_{\mu\nu}=\frac{1}{f_{\mathcal{R}}}(\mathcal{T}_{\mu\nu}^{c}
+\mathcal{T}_{\mu\nu})=\mathcal{T}_{\mu\nu}^{eff},
\end{equation}
where $\mathcal{T}_{\mu\nu}^{c}$ are the additional impacts of EMSG,
defined as
\begin{equation}\label{6}
\mathcal{T}_{\mu\nu}^{c}=\frac{1}{2}g_{\mu\nu}(f-\mathcal{R}
f_{\mathcal{R}})+(\nabla_{\mu}\nabla_{\nu}-g_{\mu\nu}\nabla
_{\mu}\nabla^{\mu})f_{\mathcal{R}}-\Theta_{\mu\nu}
f_{\mathcal{T}^{2}}.
\end{equation}
We consider static spherically spacetime to study the WH geometry as
\cite{14}
\begin{equation}\label{7}
ds^{2}=-dt^{2}e^{\alpha(r)}+
dr^{2}e^{\beta(r)}+(d\theta^{2}+d\phi^{2}\sin^{2}\theta)\eta(r),
\end{equation}
where $\eta(r)=\sinh r$, $r^2$, $\sin r$ for $K=-1,0,1$ ($K$ defines
the curvature parameter) \cite{38}. We assume
$e^{\beta(r)}=(1-\frac{h(r)}{r})^{-1}$ and $\eta(r)=r^2$ to examine
the WH structure. Here $\alpha(r)$ and $h(r)$ define the redshift
and WSF, respectively. Morris and Thorne \cite{14} stated that the
WSF must satisfy the following constraints in order to produce a
traversable WH solution
\begin{eqnarray}\label{7a}
&&h(r)-r=0 \quad  at \quad r=r_0,
\\\label{7b}
&&h'(r)<1,
\\\label{7c}
&&\frac{h(r)}{r}\rightarrow0 \quad as \quad r\rightarrow\infty,
\\\label{7d}
&&\frac{h(r)-rh(r)'}{h(r)^2}>0 \quad at \quad r=r_0.
\end{eqnarray}
Here $r_0$ is the radius of WH throat such that $r_0<r<\infty$. The
resulting field equations are
\begin{eqnarray}\nonumber
\rho^{eff}
&=&\frac{1}{f_{\mathcal{R}}}\bigg[\rho+\frac{1}{2}(\mathcal{R}
f_{\mathcal{R}}-f)+(3p^2+\rho^2+4p\rho)f_{\mathcal{T}^{2}}+
e^{-\beta} (\frac{\eta'}{\eta}-\frac{\beta'}{2})f_{\mathcal{R}}'
\\\label{8}&+&
e^{-\beta}f_{\mathcal{R}}''\bigg],
\\\label{9}
p^{eff}&=&\frac{1}{f_{\mathcal{R}}}\left[p+\frac{1}{2}(f-\mathcal
{R}f_{\mathcal{R}})-e^{-\beta}(\frac{\eta'}{\eta}+\frac{\alpha'}
{2})f_{\mathcal{R}}'\right].
\end{eqnarray}
In order to analyze the existence of some viable cosmic structures,
some constraints must be imposed on the matter named as energy
conditions. These energy bounds are classified as
\begin{itemize}
\item Null energy constraint
\begin{equation}\label{9a}
p^{eff}+\rho^{eff}\geq0.
\end{equation}
\item Strong energy constraint
\begin{eqnarray}\label{9b}
p^{eff}+\rho^{eff}\geq0, \quad 3p^{eff}+\rho^{eff}\geq0.
\end{eqnarray}
\item Dominant energy constraint
\begin{equation}\label{9c}
\rho^{eff}\pm p^{eff}\geq0.
\end{equation}
\item Weak energy constraint
\begin{eqnarray}\label{9d}
p^{eff}+\rho^{eff}\geq0,\quad \rho^{eff}\geq0.
\end{eqnarray}
\end{itemize}
These conditions must be violated for viable WH geometry. In
alternative theories of gravity, the violation of
$p^{eff}+\rho^{eff}\geq0$ demonstrates the presence of a physically
viable WH structure.

\section{Noether Symmetry Approach}

Noether symmetry offers an intriguing method for creating new cosmic
models and associated structures in alternative gravitational
theories. This method provides the first integrals of motion which
are helpful to obtain exact solutions. We use Lagrange multiplier
method as
\begin{equation}\label{10}
S=2\pi^{2}\int
\sqrt{-g}\left[f-(\mathcal{R}-\tilde{\mathcal{R}})v_{1}-
(\mathcal{T}^{2}-\mathcal{\tilde{T}}^{2})v_{2}+p\right]dr,
\end{equation}
where
\begin{eqnarray}\nonumber
\sqrt{-g}=e^{\frac{\alpha+\beta}{2}}\eta, \quad
\mathcal{\tilde{T}}^{2}=3p^{2}+\rho^{2}, \quad
v_{1}=f_{\mathcal{R}}, \quad v_{2}=f_{\mathcal{T}^{2}},
\\\label{11}
\tilde{\mathcal{R}}=-\frac{1}{e^{\beta}}\Big(\alpha''+\frac
{\alpha'^{2}}{2}+\frac{2\eta''}{\eta}+\frac{\alpha'\eta'}
{\eta}-\frac{\eta'^{2}}{2\eta^{2}}-\frac{\beta'\eta'}{\eta}
-\frac{\alpha'\beta'}{2}-\frac{2e^{\beta}}{\eta} \Big).
\end{eqnarray}
Using Eq.(\ref{11}) in (\ref{10}), we obtain
\begin{eqnarray}\nonumber
&&\mathcal{L}\left(\alpha,\beta,\eta,\mathcal{R},\mathcal{T}^{2}
,\alpha',\beta',\eta',\mathcal{R}',(\mathcal{T}^{2})'\right)= \eta
e^{\frac{\alpha+\beta}{2}}\bigg[f+p-f_{\mathcal{R}}
(\mathcal{R}-2\eta^{-1})
\\\nonumber&&
+f_{\mathcal{T}^{2}}(3p^{2}+\rho^{2}-\mathcal{T}^{2})\bigg]+\eta
e^{\frac{\alpha-\beta} {2}}\bigg[(\frac{\alpha'\eta'}{\eta}+\frac
{\eta'^{2}}{2\eta^{2}})f_{\mathcal{R}}+(\frac{2\eta'\mathcal{R}'}
{\eta}+\alpha'\mathcal{R}')f_{\mathcal{R}\mathcal{R}}
\\\label{12}
&&+(\frac{2\eta'(\mathcal{T}^{2})'}{\eta}+\alpha'
(\mathcal{T}^{2})')f_{\mathcal{R}\mathcal{T}^{2}}\bigg].
\end{eqnarray}
The Euler equations of motion and Hamiltonian of the system are
expressed as
\begin{eqnarray}\label{13}
&&\frac{\partial \mathcal{L}}{\partial q^{i}}-\frac{d}
{dr}\left(\frac{\partial \mathcal{L}}{\partial
q^{i'}}\right)=0,\quad i=1,2,3...,n
\\\label{14}&&
H=q^{i'}\left(\frac{\partial \mathcal{L}}{\partial
q^{i'}}\right)-\mathcal{L},
\end{eqnarray}
where generalized coordinates are denoted by $q^{i}$.

We use Lagrangian (\ref{12}) in Eq.(\ref{13}) and obtain
\begin{eqnarray}\nonumber
&&f-\mathcal{R}f_{\mathcal{R}}+p+f_{\mathcal{T}^{2}}(3p^{2}
+\rho^{2}+12pp_{,\alpha}+4 \rho\rho_{,\alpha}-\mathcal{T}^{2}
)+2p_{,\alpha}-\frac{1}{e^{\beta}}
\\\nonumber&&
\times\bigg[\left(\frac{2\eta''}{\eta}-\frac{\eta'^{2}}
{2\eta^{2}}-\frac{\beta'\eta'}{\eta}-\frac{2e^{\beta}}{\eta}
\right)f_{\mathcal{R}}+\left(2\mathcal{R}''-\beta'\mathcal{R}'
+\frac{2\eta'\mathcal{R}'}{\eta}\right)f_{\mathcal{R}\mathcal{R}}
\\\nonumber&&
+\left(2(\mathcal{T}^{2})''-\beta'(\mathcal{T}^{2})'+\frac{2\eta'
(\mathcal{T}^{2})'}{\eta}\right)f_{\mathcal{R}\mathcal{T}^{2}}
+2\mathcal{R}'^{2}f_{\mathcal{R}\mathcal{R}\mathcal{R}}+4\mathcal
{R}'(\mathcal{T}^{2})'f_{\mathcal{R}\mathcal{R}\mathcal{T}^{2}}
\\\label{15}&&
+2((\mathcal{T}^{2})')^{2}f_{\mathcal{R}\mathcal{T}^{2}\mathcal{T}
^{2}}\bigg]=0,
\\\nonumber&&
f-\mathcal{R}f_{\mathcal{R}}+p+f_{\mathcal{T}^{2}}(3p^{2}
+\rho^{2}+12pp_{,\beta}+4\rho\rho_{,\beta}-\mathcal{T}^{2}
)+2p_{,\beta}+\frac{1}{e^{\beta}}
\\\nonumber&&
\times\bigg[\left(\frac{2e^{\beta}}{\eta}-\frac{\eta'^{2}}
{2\eta^{2}}-\frac{\alpha'\eta'}{\eta}\right)f_{\mathcal{R}}
-\left(\alpha'\mathcal{R}'+\frac{2\eta'\mathcal{R}'}{\eta}
\right)f_{\mathcal{R}\mathcal{R}}-\alpha'(\mathcal{T}^{2})'
f_{\mathcal{R}\mathcal{T}^{2}}
\\\label{16}&&
-\frac{2\eta'(\mathcal{T}^{2})'}{\eta}f_{\mathcal{R}\mathcal{T}
^{2}}\bigg]=0,
\\\nonumber&&
f-\mathcal{R}f_{\mathcal{R}}+f_{\mathcal{T}^{2}}(3p^{2}
+\rho^{2}+6\eta
pp_{,\eta}+2\eta\rho\rho_{,\eta}-\mathcal{T}^{2})-\frac{1}
{e^{\beta}}\bigg[\big(\alpha''+\frac{\alpha'^{2}}{2}
\\\nonumber&&
+\frac{\eta''}{\eta}+\frac{\alpha'\eta'}{2\eta}-\frac{\beta'\eta'}
{2\eta}-\frac{\alpha'\beta'}{2}-\frac{\eta'^{2}}{2\eta^{2}}\big)
f_{\mathcal{R}}+2\mathcal{R}'^{2}f_{\mathcal{R}\mathcal{R}\mathcal
{R}}+4\mathcal{R}'(\mathcal{T}^{2})'f_{\mathcal{R}\mathcal{R}
\mathcal{T}^{2}}
\\\nonumber&&
+\left(\alpha'\mathcal{R}'-\beta'\mathcal{R}'+2\mathcal{R}''
+\frac{\eta'\mathcal{R}'}{\eta}\right)f_{\mathcal{R}\mathcal{R}}
+\left(\alpha'(\mathcal{T}^{2})'-\beta'(\mathcal{T}^{2})
+2(\mathcal{T}^{2})'' \right.\\\label{17}&&\left.
+\frac{\eta'(\mathcal{T}^{2})'}{\eta}\right)f_{\mathcal{R}
\mathcal{T}^{2}}+2((\mathcal{T}^{2})')^{2}f_{\mathcal{R}
\mathcal{T}^{2}\mathcal{T}^{2}}-pe^{\beta} -\eta p_{,\eta}
e^{\beta}\bigg]=0,
\\\nonumber&&
+e^{\beta}(\mathcal{R}-2\eta^{-1})f_{\mathcal{R}\mathcal{R}}
-e^{\beta}(3p^{2}+\rho^{2}-\mathcal{T}^{2})f_{\mathcal{R}
\mathcal{T}^{2}}+\bigg[\alpha''+\frac{\alpha'^{2}}{2}
+\frac{2\eta''}{\eta}
\\\label{18}&&
+\frac{\alpha'\eta'}{\eta}-\frac{\beta'\eta'}{\eta}-\frac
{\alpha'\beta'}{2}-\frac{\eta'^{2}}{2\eta^{2}}\bigg]
f_{\mathcal{R}\mathcal{R}}=0,
\\\nonumber&&
e^{\beta}(\mathcal{R}-2\eta^{-1})f_{\mathcal{R}\mathcal{T}^{2}}
-e^{\beta}(3p^{2}+\rho^{2}-\mathcal{T}^{2})f_{\mathcal{T}^{2}
\mathcal{T}^{2}}+\bigg[\alpha''+\frac{\alpha'^{2}}{2}
+\frac{2\eta''}{\eta}
\\\label{19}&&
+\frac{\alpha'\eta'}{\eta}-\frac{\beta'\eta'}{\eta}-\frac{\alpha'
\beta'}{2}-\frac{\eta'^{2}}{2\eta^{2}}\bigg]f_{\mathcal{R}
\mathcal{T}^{2}}=0.
\end{eqnarray}
Using Eq.(\ref{12}) in (\ref{14}), it follows
\begin{equation}\label{20}
e^{\beta(r)} =\frac{\left(\frac{\eta'^{2}}{2\eta^{2}}
+\frac{\alpha'\eta'}{\eta}\right)f_{\mathcal{R}}+\left
(\alpha'+\frac{2\mathrm{\eta}'}{\eta}\right)\Big
(\mathcal{R}'f_{\mathcal{R}\mathcal{R}}+(\mathcal
{T}^{2})'f_{\mathcal{T}^{2}\mathcal{T}^{2}}\Big)}
{\Big(f-\mathcal{R}f_{\mathcal{R}}+\left(3p^{2}
+\rho^{2}-\mathcal{T}^{2}\right)f_{\mathcal{T}^{2}}
+p+\frac{2f_{\mathcal{R}}}{\eta}\Big)}.
\end{equation}
The symmetry generators are considered as
\begin{equation}\label{21}
\mathcal{K}= \lambda\frac{\partial}{\partial r}
+\gamma^{i}\frac{\partial}{\partial q^{i}},
\quad i= 1,2,3,4,5.
\end{equation}
where $\lambda=
\lambda(\alpha,\beta,\eta,\mathcal{R},\mathcal{T}^{2})$ and
$\gamma=\gamma(\alpha,\beta,\eta,\mathcal{R},\mathcal{T}^{2})$ are
unknown coefficients of the vector field. For the existence of NS,
the Lagrangian must satisfy the following invariance constraint
\begin{equation}\label{22}
\mathcal{K}^{[1]}\mathcal{L}+(D\lambda)\mathcal{L}= D\Psi,
\end{equation}
where total derivative, prolongation of first order and boundary
term are represented by $D$, $\mathcal{K}^{[1]}$ and $\Psi$,
respectively. Further, it is determined as
\begin{equation}\label{23}
\mathcal{K}^{[1]}= \mathcal{K}+{\gamma^{i}}'\frac{\partial}
{\partial{q^{i}}'}, ~~~D=\frac{\partial}{\partial
r}+{q^{i}}'\frac{\partial} {\partial {q^{i}}}.
\end{equation}
Here, ${\gamma^{i}}'= D{\gamma^{i}}'-{q^{i}}'D\lambda$. The
conserved quantities are expressed as
\begin{equation}\label{24}
I= -\lambda H+ \gamma^{i}\frac{\partial \mathcal{L}}{\partial
q^{i}}-\Psi,
\end{equation}
which play an important role for developing the viable solutions.
The coefficients of Eq.(\ref{22}) are given in Appendix \textbf{A}.

\section{Exact Solutions}

This section formulates the symmetry generators, conserved
quantities and the corresponding viable solutions using the above
system of PDEs. The system's complexity decreases via the NS
technique, which also helps to find the exact solutions. Thus, it
would be interesting to study viable and traversable WH solutions
using this approach. However, the aforementioned system is more
complex, so it is very difficult to find exact solutions without
taking the EMSG model. We assume the EMSG model as \cite{39}
\begin{equation}\label{45}
\mathcal{R}+\chi\mathcal{T}^{2}=f(\mathcal{R},\mathcal{T}^{2}).
\end{equation}
where we take constant $\chi=1$ for our convenience. We include
cosmological constant in this model to make the resemblance with the
standard $\Lambda$CDM model as
\begin{equation}\label{46}
\mathcal{R}+\Lambda(\mathcal{T}^{2})+\mathcal{T}^{2}=f(\mathcal{R}
,\mathcal{T}^{2}).
\end{equation}
The exact solutions of the system of equations (\ref{25})-(\ref{43})
are
\begin{eqnarray}\nonumber
&&\gamma^{2}=-\frac{2\xi_{2}\xi_{5}}{r^{2}}, \quad \lambda=
\xi_{1}-\frac {\xi_{2}\xi_{5}}{r}, \quad
\gamma^{1}=\gamma^{3}=\gamma^{4}=\gamma^{5}=0,\\\label{47}
&&\Lambda(\mathcal{T}^{2})=-\mathcal{T}^{2}
+\xi_{3}\mathcal{T}^{2}+\xi_{4}, \quad \Psi= \xi_{5}r,
\end{eqnarray}
where arbitrary constants are denoted by $\xi_{i}$. It is important
to consider isotropic matter since it accurately explains the
composition of matter in various celestial objects. Dust fluid can
also analyze the configuration of matter only in the presence of
negligible amount of radiation. Here, we examine the existence of
viable traversable WH structures for dust and non-dust fluid
configurations.

\subsection{Dust Case}

Equation (\ref{4}) becomes
\begin{equation}\label{48}
\mathcal{T}_{\mu\nu}=\rho\mathcal{U}_{\mu}\mathcal{U}_{\nu}.
\end{equation}
Using Eq.(\ref{48}) in (\ref{44}), we have
\begin{eqnarray}\label{49}
\rho= \sqrt{\frac{e^\frac{-\alpha-\beta}{2}}{2\xi_{2} \xi_{3}}},
\quad
\mathcal{R}+2\xi_{3}\mathcal{T}^{2}+\xi_{4}=f(\mathcal{R},
\mathcal{T}^{2}).
\end{eqnarray}
The NS generators and corresponding first integrals of motion become
\begin{eqnarray}\label{49a}
\mathcal{K}_{1}&=&\frac{\partial}{\partial r}, \quad
\mathcal{K}_{2}=-\frac{2\xi_{2}}{r} \frac{\partial} {\partial
r}-\frac{2\xi_{2}}{r^{2}}\frac{\partial} {\partial \beta},
\\\label{49b}
I_{1}&=&2e^{\frac{\alpha-\beta}{2}}\bigg[1+\alpha'
r-\left(1+\frac{\xi_{4}r^{2}}{2}+\frac{r^{2}e^{\frac
{-\alpha-\beta}{2}}}{2\xi_{2}}\right)e^{\beta}\bigg],
\\\label{49c}
I_{2}&=&r-\frac{2\xi_{2}e^{\frac{\alpha-\beta}{2}}}{r}
\bigg[1+\alpha'r-\left(1+\frac{\xi_{4}r^{2}}{2}+\frac
{r^{2}e^{\frac{-\alpha-\beta}{2}}}{2\xi_{2}}\right) e^{\beta}\bigg].
\end{eqnarray}
Substituting Eq.(\ref{49}) in (\ref{20}), we obtain
\begin{equation}\label{50}
e^{\beta(r)} = \frac{1+\alpha'r}{1+\frac{r^{2}\xi_{4}}
{2}+\frac{r^{2}e^{\frac{-\alpha-\beta}{2}}}{2\xi_{2}}}.
\end{equation}
We consider different redshift functions as
\begin{eqnarray}\label{50a}
\alpha(r)=j\ln(\frac{r}{r_{0}})~~ \cite{40}, \quad
\alpha(r)=e^{-\frac{r_{0}}{r}}~~ \cite{41},
\end{eqnarray}
to examine the viable WH geometry through null energy condition and
WSF. Here, $j$ is an arbitrary constant. We manipulate Eq.(\ref{50})
for the considered redshift functions in the following cases.

\subsubsection*{Case I: $\alpha(r)=j\ln(\frac{r}{r_{0}})$}

Substituting this value in (\ref{50}), it follows
\begin{eqnarray}\nonumber
\beta(r)&=&4\ln(2)-2\ln\bigg[\frac{1}{\xi_{2}(r+j)}\bigg\{r^{2}
+\bigg\{8(\frac{r}{r_{0}})^{j}\xi_{2}^{2}\xi_{4}jr^{2}
+8(\frac{r}{r_{0}})^{j}\xi_{2}^{2}\xi_{4}r^{3}
\\\label{50b}
&+&16(\frac{r}{r_{0}})^{j}\xi_{2}^{2}j+16(\frac{r}{r_{0}})
^{j}\xi_{2}^{2}r+r^{4}\bigg\}^{\frac{1}{2}}\bigg\}(\frac{r}
{r_{0}})^{-j/2}\bigg].
\end{eqnarray}
The corresponding WSF is
\begin{eqnarray}\nonumber
h(r)&=&-\frac{r}{8\xi_{2}^{2}(r+j)^{2}}\bigg[4\xi_{2}^{2}
\xi_{4}jr^{2}+4\xi_{2}^{2}\xi_{4}r^{3}+(\frac{r}{r_{0}})
^{-j}r^{4}+(\frac{r}{r_{0}})^{-j}\bigg\{8jr^{2}\xi_{2}^{2}
\\\nonumber&\times&
(\frac{r}{r_{0}})^{j}\xi_{4}+8(\frac{r}{r_{0}})^{j}\xi_{2}
^{2}\xi_{4}r^{3}+16(\frac{r}{r_{0}})^{j}\xi_{2}^{2}j+16
(\frac{r}{r_{0}})^{j}\xi_{2}^{2}r+r^{4}r^{2}\bigg\}
^{\frac{1}{2}}
\\\label{50c}&-&
8\xi_{2}^{2}j^{2}-16\xi_{2}^{2}jr-8\xi_{2}^{2}r^{2}
+8\xi_{2}^{2}j+8\xi_{2}^{2}r\bigg].
\end{eqnarray}
The energy density becomes
\begin{eqnarray}\nonumber
\rho&=&\frac{\sqrt{2}}{4}\bigg[\frac{1}{\xi_{2}^{2}(r+j)
\xi_{3}}\bigg\{{r}^{2}+\bigg\{8(\frac{r}{r_{0}})^{j}\xi
_{2}^{2}\xi_{4}jr^{2}+8(\frac{r}{r_{0}})^{j}\xi_{2}^{2}
\xi_{4}r^{3}+16(\frac{r}{r_{0}})^{j}\xi_{2}^{2}j
\\\label{50d}
&+&16(\frac{r}{r_{0}})^{j}\xi_{2}^{2}r+r^{4}\bigg\}
^\frac{1}{2}\bigg\}(\frac{r}{r_{0}})^{-j}\bigg]
^{\frac{1}{2}}.
\end{eqnarray}
The null energy condition turns out to be
\begin{eqnarray}\nonumber
\rho^{eff}+p^{eff}&=&\frac{\sqrt{2}}{4}\bigg[\frac{1}
{\xi_{2}^{2}(r+j)\xi_{3}}\bigg\{{r}^{2}+\bigg\{8(\frac
{r}{r_{0}})^{j}\xi_{2}^{2}\xi_{4}jr^{2}+8(\frac{r}
{r_{0}})^{j}\xi_{2}^{2}\xi_{4}r^{3}+16j
\\\nonumber
&\times&(\frac{r}{r_{0}})^{j}\xi_{2}^{2}+16(\frac{r}
{r_{0}})^{j}\xi_{2}^{2}r+r^{4}\bigg\}^\frac{1}{2}
\bigg\}(\frac{r}{r_{0}})^{-j}\bigg]^{\frac{1}{2}}
+\frac{1}{4}\bigg[\frac{1}{\xi_{2}^{2}(r+j)}\bigg\{
{r}^{2}
\\\nonumber
&+&\bigg\{8(\frac{r}{r_{0}})^{j}\xi_{2}^{2}\xi_{4}j
r^{2}+8(\frac{r}{r_{0}})^{j}\xi_{2}^{2}\xi_{4}r^{3}
+16(\frac{r}{r_{0}})^{j}\xi_{2}^{2}j+16(\frac{r}
{r_{0}})^{j}\xi_{2}^{2}r
\\\label{50e}
&+&r^{4}\bigg\}^\frac{1}{2} \bigg\}(\frac {r}{r_{0}})
^{-j}\bigg].
\end{eqnarray}
We investigate the graphical behavior of WSF in the Figure
\textbf{1}. In the upper panel, the left graph shows that the
behavior of WSF is positive with $h(r)<r$ whereas the right graph
represents asymptotically flat behavior, i.e., $h(r)\rightarrow0$
when $r\rightarrow\infty$. The WH throat exists at $r_{0}=0.001$ and
$\frac{dh(r_{0})}{dr}<1$ as shown in the below panel of left and
right graphs, respectively The last plot shows that flaring-out
condition is satisfied at wormhole throat. The graphical behavior of
the null energy condition is given in Figure \textbf{2}, which shows
that the effective fluid variables violate null energy condition $(
\rho^{eff}+p^{eff}\leq0$), hence ensures the presence of traversable
WH geometry.
\begin{figure}
\epsfig{file=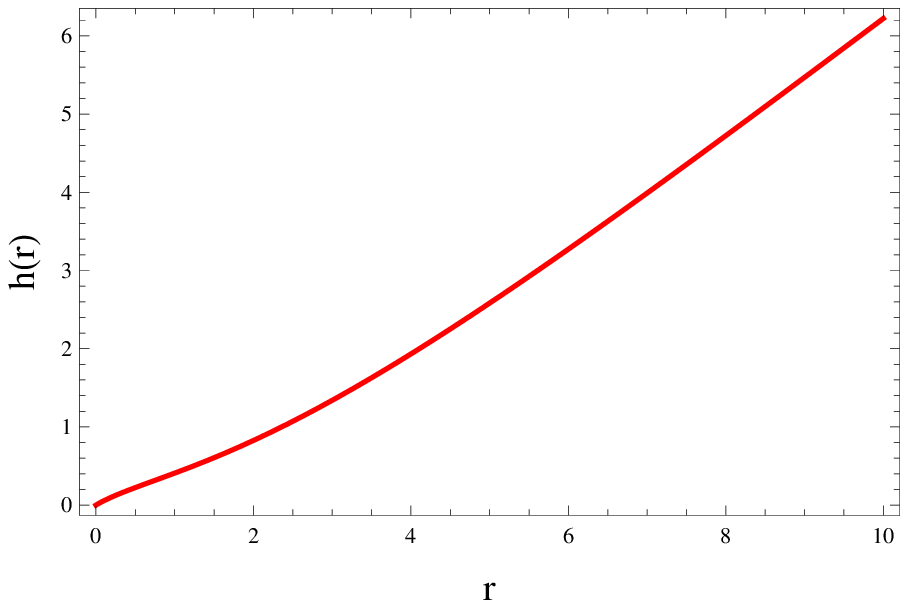,width=.5\linewidth}
\epsfig{file=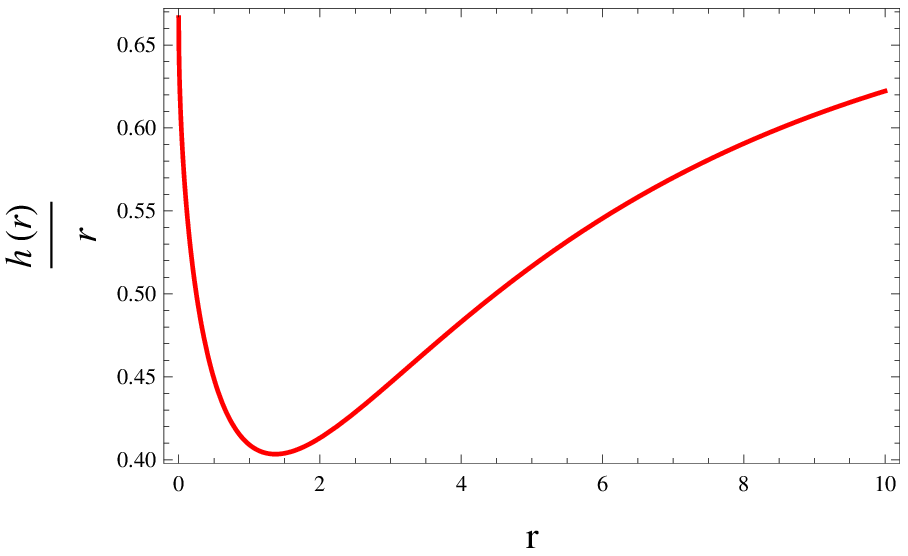,width=.5\linewidth}
\epsfig{file=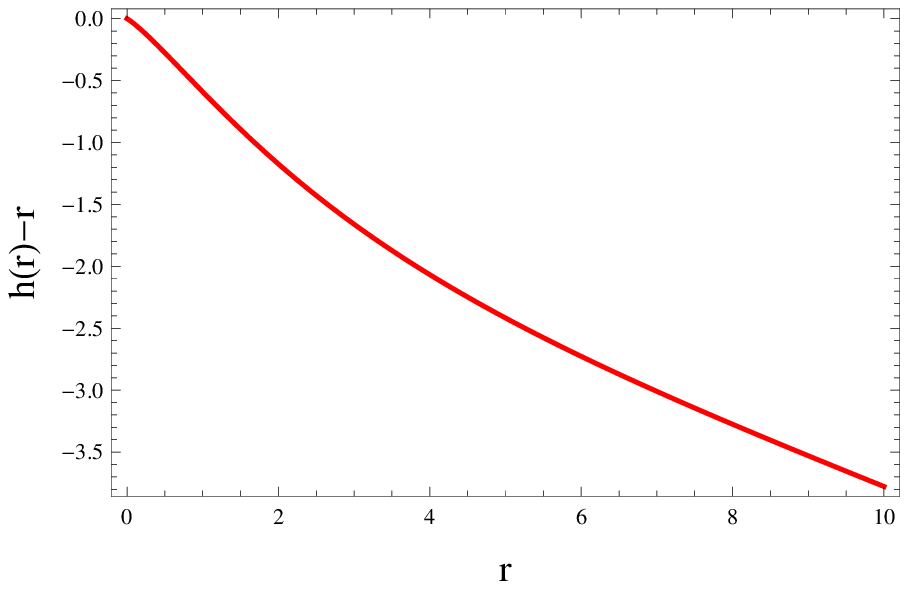,width=.5\linewidth}
\epsfig{file=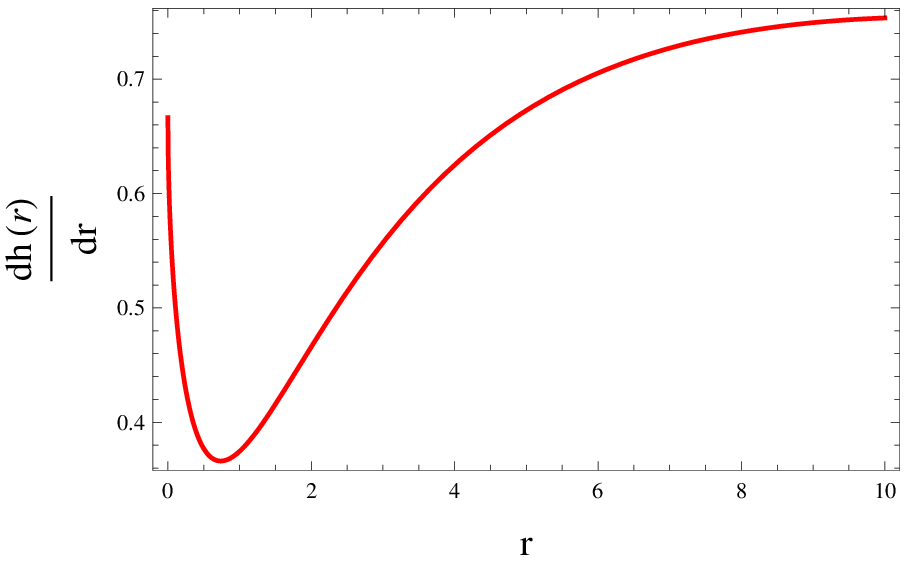,width=.5\linewidth}\center
\epsfig{file=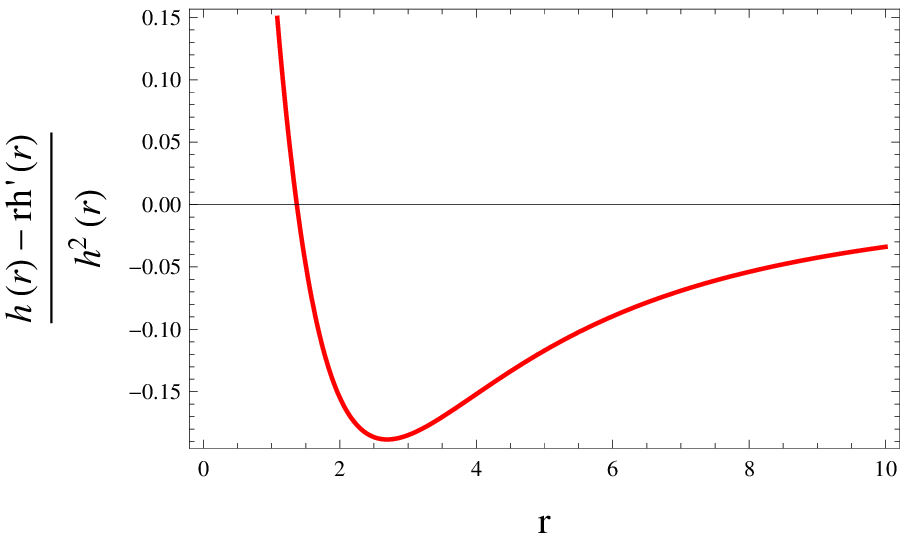,width=.5\linewidth}\caption{Graphs of WSF
versus $r$.}
\end{figure}
\begin{figure}\center
\epsfig{file=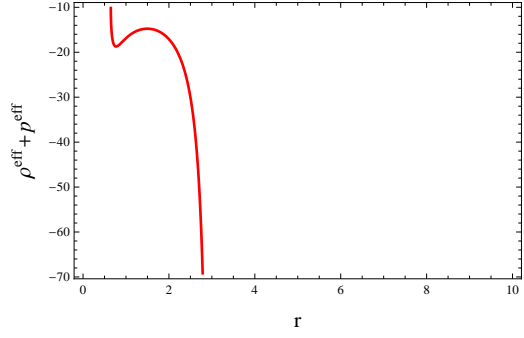,width=.5\linewidth}\caption{Behavior of null
energy condition versus $r$.}
\end{figure}

\subsubsection*{Case II: $\alpha(r)=e^{\frac{r_{0}}{r}}$}

Here, Eq.(\ref{50}) gives
\begin{eqnarray}\nonumber
\beta(r)&=&2\ln\bigg[\frac{1}{4\xi_{2}}\bigg\{r^{3}+\bigg
\{8(e^{\frac{1}{2}(e^{\frac{r_{0}}{r}})^{-1}})^{2}\xi_{2}
^{2}\xi_{4}r^{5}+8e^{-\frac{1}{2}\frac{1}{r}(2r_{0}
e^{\frac{r_{0}}{r}}-r)(e^{\frac{r_{0}}{r}})^{-1}}e^{\frac
{1}{2}(e^{\frac{r_{0}}{r}})^{-1}}
\\\nonumber&\times&
\xi_{2}^{2}\xi_{4}r_{0}r^{3}+16(e^{\frac{1}{2}(e^\frac
{r_{0}}{r})^{-1}})^{2}\xi_{2}^{2}r^{3}+16e^{-\frac{1}{2}
\frac{1}{r}(2r_{0}e^{\frac{r_{0}}{r}}-r)(e^{\frac{r_{0}}
{r}})^{-1}}e^{\frac{1}{2}(e^{\frac{r_{0}}{r}})^{-1}}
\xi_{2}^{2}r_{0}r
\\\label{51}&+&
r^{6}\bigg\}^\frac{1}{2}\bigg\}\bigg\{e^{\frac{1}{2}
(e^{\frac{r_{0}}{r}})^{-1}}r^{2}+e^{-\frac{1}{2}\frac{1}
{r}(2r_{0}e^{\frac{r_{0}}{r}}-r)(e^{\frac{r_{0}}{r}})
^{-1}}r_{0}\bigg\}^{-1}\bigg].
\end{eqnarray}
The corresponding WSF becomes
\begin{eqnarray}\nonumber
b(r)&=&\bigg[1-16\xi_{2}^{2}\bigg\{e^{\frac{1}{2}(e^{\frac
{r_{0}}{r}})^{-1}}r^{2}+e^{-\frac{1}{2}\frac{1}{r}(2r_{0}
e^{\frac{r_{0}}{r}}-r)(e^{\frac{r_{0}}{r}})^{-1}}r_{0}
\bigg\}^{2}\bigg[r^{3}+\bigg\{(e^{\frac{1}{2}(e^{\frac{
r_{0}}{r}})^{-1}})^{2}
\\\nonumber
&\times&8\xi_{2}^{2}\xi_{4}r^{5}+8e^{-\frac{1}{2}{\frac
{1}{r}(2r_{0}e^{\frac{r_{0}}{r}}-r)(e^{\frac{r_{0}}{r}})
^{-1}}}e^{\frac{1}{2}(e^{\frac{r_{0}}{r}})^{-1}}\xi_{2}
^{2}\xi_{4}r_{0}r^{3}+16(e^{\frac{1}{2}(e^{\frac{r_{0}}
{r}})^{-1}})^{2}\xi_{2}^{2}r^{3}
\\\label{52}
&+&16e^{-\frac{1}{2}\frac{1}{r}(2r_{0}e^{\frac{r_{0}}{r}
}-r)(e^{\frac{r_{0}}{r}})^{-1}}e^{\frac{1}{2}(e^{\frac
{r_{0}}{r}})^{-1}}\xi_{2}^{2}r_{0}r+r^{6}\bigg\}^{\frac
{1}{2}}\bigg]^{-2}\bigg]r.
\end{eqnarray}
The energy density in this case is given as
\begin{eqnarray}\nonumber
\rho&=&\sqrt{2}\bigg[\frac{e^{-\frac{1}{2}e^{-\frac{r_{0}}
{r}}}}{\xi_{3}}\bigg\{e^{1/2e^{-\frac{r_{0}}{r}}}r^{2}
+e^{-1/2\frac{1}{r}(2r_{0}e^{\frac{r_{0}}{r}}-r)e^{-\frac
{r_{0}}{r}}}r_{0}\bigg\}\bigg[r^{3}+\bigg\{r\bigg(8
e^{e^{-\frac{r_{0}}{r}}}\xi_{2}^{2}\xi_{4}r^{4}
\\\nonumber
&+&8e^{-\frac{1}{r}e^{-\frac{r_{0}}{r}}(r_{0}e^{\frac
{r_{0}}{r}}-r)}\xi_{2}^{2}\xi_{4}r_{0}r^{2}+16e^{e^{
-\frac{r_{0}}{r}}}\xi_{2}^{2}r^{2}+r^{5}+e^{-\frac{1}
{r}e^{-\frac{r_{0}}{r}}}(r_{0}e^{\frac{r_{0}}{r}}-r)
\\\label{53}
&\times&16\xi_{2}^{2}r_{0}\bigg)
\bigg\}^{\frac{1}{2}}\bigg]^{-1}\bigg]^{\frac{1}{2}}.
\end{eqnarray}
Substituting the value of redshift function and $\beta(r)$ in
Eqs.(\ref{8}) and (\ref{9}), we have
\begin{eqnarray}\nonumber
\rho^{eff}+p^{eff}&=&\sqrt{2}\bigg[\frac{e^{-\frac{1}{2}
e^{-\frac{r_{0}}{r}}}}{\xi_{3}}\bigg\{e^{\frac{1}{2}e^{-
\frac{r_{0}}{r}}}r^{2}+e^{-\frac{1}{2}\frac{1}{r}(2r_{0}
e^{\frac{r_{0}}{r}}-r)e^{-\frac{r_{0}}{r}}}r_{0}\bigg\}
\bigg[r^{3}+\bigg\{r\bigg(8r^{4}
\\\nonumber
&\times&e^{e^{-\frac{r_{0}}{r}}}\xi_{2}^{2}\xi_{4}+8e^{
-\frac{1}{r}e^{-\frac{r_{0}}{r}}(r_{0}e^{\frac{r_{0}}{r}
}-r)}\xi_{2}^{2}\xi_{4}r_{0}r^{2}+16e^{e^{-\frac{r_{0}}
{r}}}\xi_{2}^{2}r^{2}+r^{5}+\xi_{2}^{2}r_{0}
\\\nonumber
&\times&16e^{-\frac{1}{r}e^{-\frac{r_{0}}{r}}}(r_{0}
e^{\frac{r_{0}}{r}}-r)\bigg)\bigg\}^{\frac{1}{2}}\bigg]
^{-1}\bigg]^{\frac{1}{2}}+\xi_{3}\bigg[\frac{e^{-1/2e
^{-\frac{r_{0}}{r}}}}{\xi_{3}}\bigg\{e^{\frac{1}{2}e
^{-\frac{r_{0}}{r}}}r^{2}
\\\nonumber
&+&e^{-1/2\frac{1}{r}(2r_{0}e^{\frac{r_{0}}{r}}-r)e^
{-\frac{r_{0}}{r}}}r_{0}\bigg\}\bigg[r^{3}+\bigg\{r
\bigg(8e^{e^{-\frac{r_{0}}{r}}}\xi_{2}^{2}\xi_{4}r^{4}
+e^{-\frac{1}{r}e^{-\frac{r_{0}}{r}}(r_{0}e^{\frac{
r_{0}}{r}}-r)}
\\\nonumber
&\times&8\xi_{2}^{2}
\xi_{4}r_{0}r^{2}+16e^{e^{-\frac{r_{0}}{r}}}\xi_{2}
^{2}r^{2}+r^{5}+e^{-\frac{1}{r}e^{-\frac{r_{0}}{r}}}
(r_{0}e^{\frac{r_{0}}{r}}-r)
\\\label{54}
&\times&16\xi_{2}^{2}r_{0}\bigg)\bigg\}^{\frac{1}{2}}
\bigg]^{-1}\bigg].
\end{eqnarray}
In Figure \textbf{3}, the upper right plot determines that the
behavior of shape function is not asymptotically flat while left
plot implies that $h(r)< r$. The right plot in the below panel shows
$\frac{dh(r_{0})}{dr}<1$ and the left plot determines that WH throat
exist at $r_{0}=0.01$.  Also, the flaring-out condition satisfies at
wormhole throat. Figure \textbf{4} violates the null energy
condition which manifests the existence of viable traversable WH
geometry.
\begin{figure}
\epsfig{file=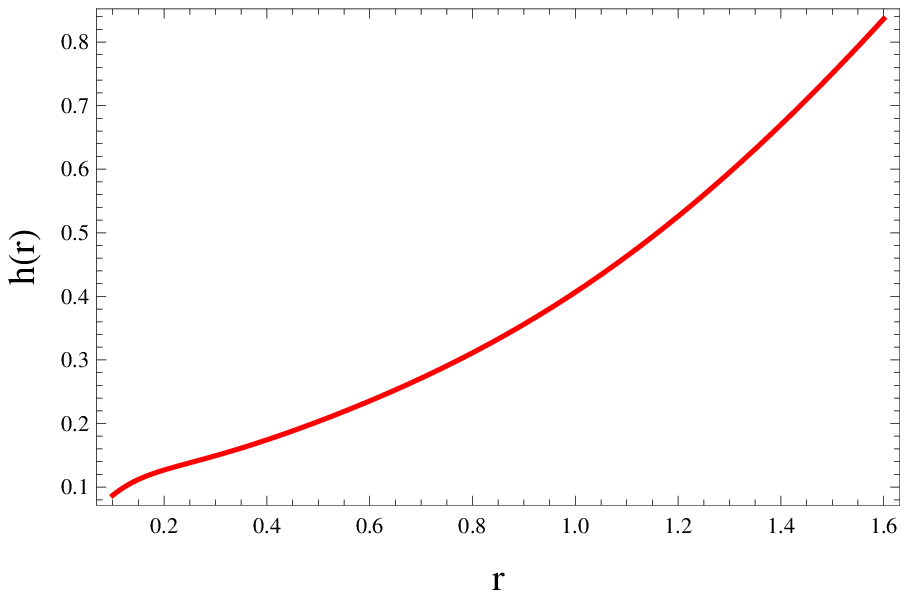,width=.5\linewidth}
\epsfig{file=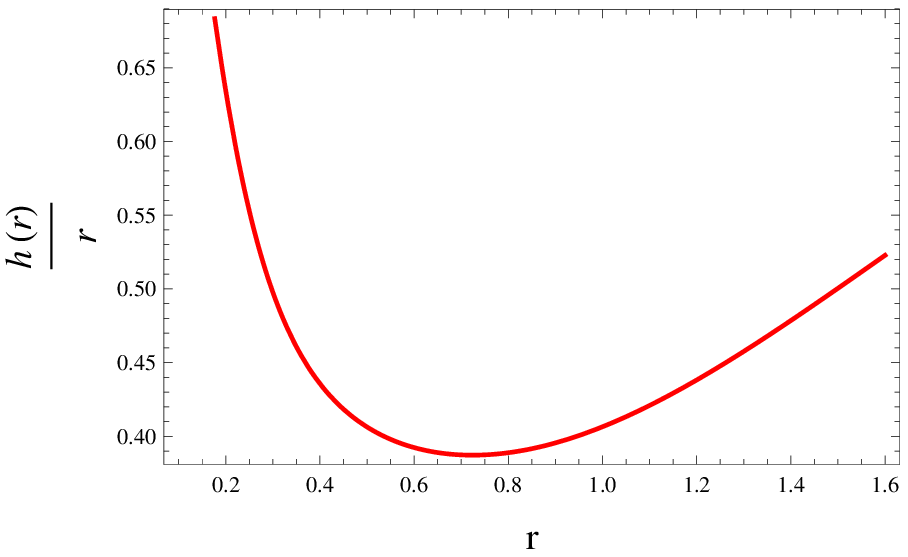,width=.5\linewidth}
\epsfig{file=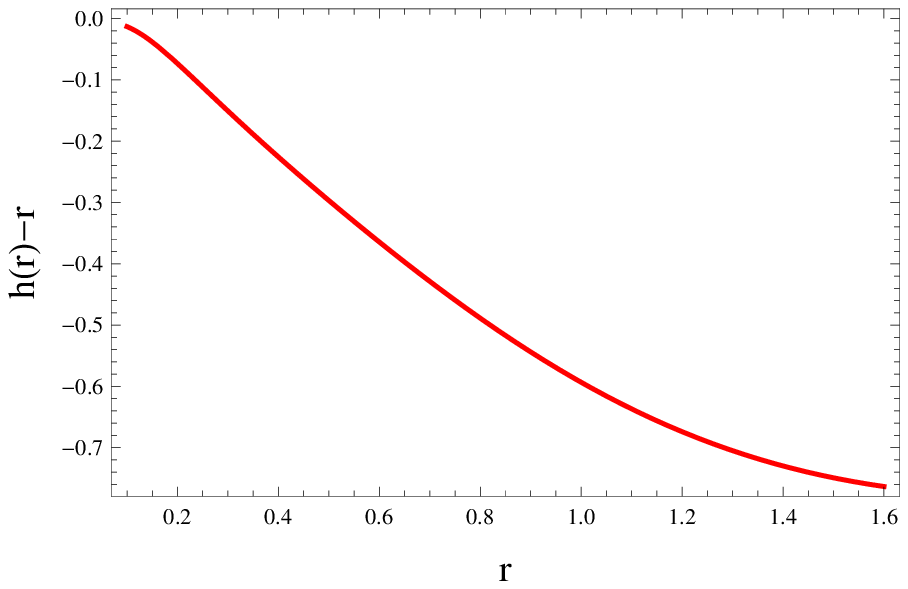,width=.5\linewidth}
\epsfig{file=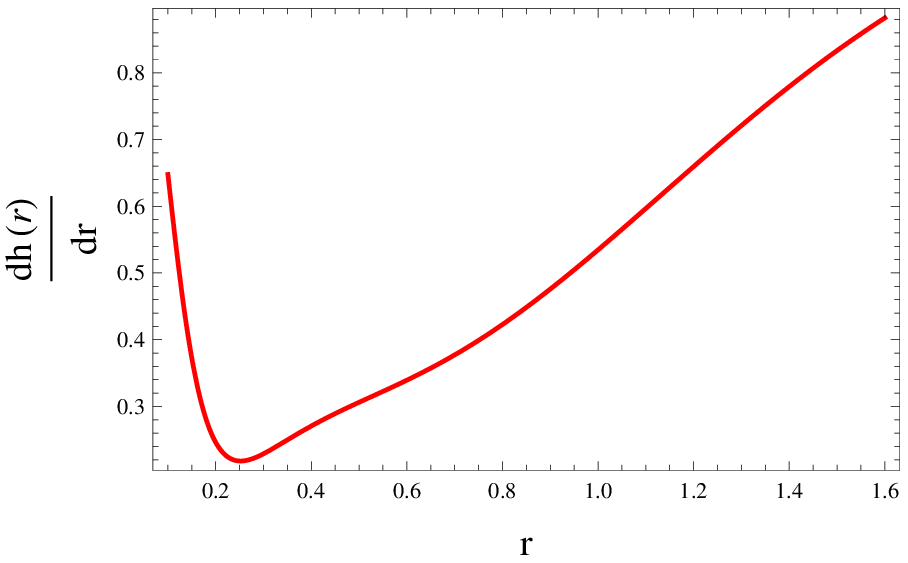,width=.5\linewidth}\center
\epsfig{file=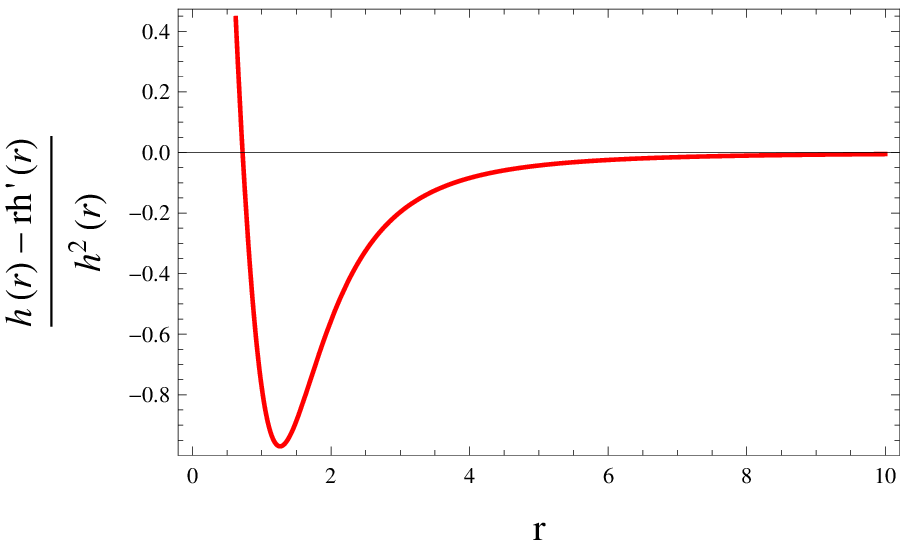,width=.5\linewidth} \caption{Graphs of WSF
corresponding to radial coordinate.}
\end{figure}
\begin{figure}\center
\epsfig{file=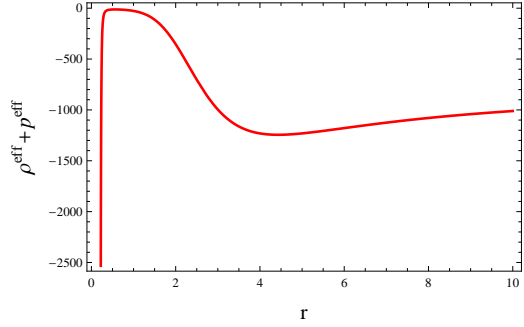,width=.5\linewidth}\caption{Graphs of null
energy condition versus $r$.}
\end{figure}

\subsection{Non-Dust Case}

Here, we assume a specific relation between fluid parameters as
$p=\omega\rho$ ($\omega$ is EoS parameter) and manipulate
Eq.(\ref{44}) which gives
\begin{equation}\label{55}
\rho=\frac{-\xi_{2}\omega+\sqrt{\xi_{2}^2\omega^2+4\xi_{2}\xi_{3}
e^\frac{-\alpha-\beta}{2}+12\xi_{2}\xi_{3}\omega^2e^\frac{-\alpha
-\beta}{2}}}{2\xi_{2}\xi_{3}(3\omega^2+1)}.
\end{equation}
The NS generators and the corresponding integral of motion yield
\begin{eqnarray}\label{55a}
\mathcal{K}_{1}&=&\frac{\partial}{\partial r}, \quad
\mathcal{K}_{2}=-\frac{2\xi_{2}}{r} \frac{\partial}{\partial
r}-\frac{2\xi_{2}}{r^{2}}\frac{\partial} {\partial \beta},
\\\nonumber
I_{1}&=&e^{\frac{\alpha-\beta}{2}}r^{2}\bigg[2(\frac{1
+\alpha'r}{r^2})-e^{\beta}\left(\xi_{4}+\frac{2}{r^2}
+\left(\xi_{3}\left(3\omega^2+1\right)\right)
\right.\\\nonumber&\times&\left.
\left(\frac{-\xi_{2}\omega+\sqrt{\xi_{2}^2\omega^2+4
\xi_{2}\xi_{3}e^\frac{-\alpha-\beta}{2}+12\xi_{2}
\xi_{3}\omega^2e^\frac{-\alpha-\beta}{2}}}{2\xi_{2}
\xi_{3}\left(3\omega^2+1\right)}\right)^{2}
\right.\\\label{55b}&+&\left.
\omega\frac{-\xi_{2}\omega+\sqrt{\xi_{2}^2\omega^2+4
\xi_{2}\xi_{3}e^\frac{-\alpha-\beta}{2}+12\xi_{2}
\xi_{3}\omega^2e^\frac{-\alpha-\beta}{2}}}{2\xi_{2}
\xi_{3}\left(3\omega^2+1\right)}\right) \bigg],
\\\nonumber
I_{2} &=& r-\xi_{2}re^{\frac{\alpha-\beta}{2}}\bigg
[\frac{2\alpha'}{r}+\frac{2}{r^2}-e^{\beta}\left
(\xi_{4}+\frac{2}{r^2}+\left(\xi_{3}\left(3\omega^2 +1\right)\right)
\right.\\\nonumber&\times&\left.
\left(\frac{-\xi_{2}\omega+\sqrt{\xi_{2}^2\omega^2
+4\xi_{2}\xi_{3}e^\frac{-\alpha-\beta}{2}+12\xi_{2}
\xi_{3}\omega^2e^\frac{-\alpha-\beta}{2}}}{2\xi_{2}
\xi_{3}\left(3\omega^2+1\right)}\right)^{2}
\right.\\\label{55c}&+&\left.
\omega\frac{-\xi_{2}\omega+\sqrt{\xi_{2}^2\omega^2
+4\xi_{2}\xi_{3}e^\frac{-\alpha-\beta}{2}+12\xi_{2}
\xi_{3}\omega^2e^\frac{-\alpha-\beta}{2}}}{2\xi_{2}
\xi_{3}\left(3\omega^2+1\right)}\right) \bigg].
\end{eqnarray}
Using Eq.(\ref{55}) in (\ref{20}), we have
\begin{equation}\label{56}
e^{\beta(r)}= \frac{2+2\alpha'r}{r^{2}\left(\xi_{4}
+\frac{2}{r^2}+\frac{e^{\frac{-\alpha-\beta}{2}}}
{\xi_{2}} \right)}.
\end{equation}
Now, we examine the presence of viable WH geometry for the identical
redshift functions that were studied in the dust case.

\subsubsection*{Case I: $\alpha(r)=j\ln(\frac{r}{r_{0}})$}

Equation (\ref{56}) in this case turns out to be
\begin{eqnarray}\nonumber
\beta(r)&=&4\ln(2)-2\ln\bigg[\frac{1}{\xi_{2}(r+j)}\bigg
\{r^{2}+\bigg\{8(\frac{r}{r_{0}})^{j}\xi_{2}^{2}\xi_{4}j
r^{2}+8(\frac{r}{r_{0}})^{j}\xi_{2}^{2}\xi_{4}r^{3}
\\\label{57}
&+&16j(\frac{r}{r_{0}})^{j}\xi_{2}^{2}+16(\frac{r}{r_{0}})
^{j}\xi_{2}^{2}r+r^{4}\bigg\}^{\frac{1}{2}}\bigg\}(\frac
{r}{r_{0}})^{-\frac{j}{2}}\bigg].
\end{eqnarray}
The WSF becomes
\begin{eqnarray}\nonumber
b(r)&=&-\frac{r}{8\xi_{2}^{2}(r+j)^{2}}\bigg[4\xi_{2}^{2}
\xi_{4}jr^{2}+4\xi_{2}^{2}\xi_{4}r^{3}+(\frac{r}{r_{0}})
^{-j}r^{4}+(\frac{r}{r_{0}})^{-j}\bigg\{8(\frac{r}{r_{0}
})^{j}
\\\nonumber&\times&
\xi_{2}^{2}\xi_{4}jr^{2}+8(\frac{r}{r_{0}})^{j}\xi_{2}^{2}
\xi_{4}r^{3}+16(\frac{r}{r_{0}})^{j}\xi_{2}^{2}j+16(\frac
{r}{r_{0}})^{j}\xi_{2}^{2}r+r^{4}r^{2}\bigg\}^{\frac{1}{2}}
\\\label{58}&-&
8\xi_{2}^{2}j^{2}-16\xi_{2}^{2}jr-8\xi_{2}^{2}r^{2}+8\xi_{2}
^{2}j+8\xi_{2}^{2}r\bigg].
\end{eqnarray}
Substituting Eq.(\ref{57}) in (\ref{55}), we obtain
\begin{eqnarray}\nonumber
\rho&=&\frac{1}{2\xi_{2}\xi_{3}(3\omega^{2}+1)}\bigg[-\xi_{2}
\omega+\bigg\{\frac{1}{r+j}\bigg(3\omega^{2}\xi_{3}(\frac{r}
{r_{0}})^{-j}r^{2}+3(\frac{r}{r_{0}})^{-j}\bigg(8(\frac{r}
{r_{0}})^{j}
\\\nonumber
&\times&\xi_{2}^{2}c_4jr^{2}+8(\frac{r}{r_{0}})^{j}\xi_{2}^{2}
c_4r^{3}+16(\frac{r}{r_{0}})^{j}\xi_{2}^{2}j+16(\frac{r}{r_{0}
})^{j}\xi_{2}^{2}r+r^{4}\bigg)^{\frac{1}{2}}\xi_{3}\omega^{2}
\\\nonumber
&+&\xi_{3}(\frac{r}{r_{0}})^{-j}r^{2}+\xi_{2}^{2}\omega^{2}j
+\xi_{2}^{2}\omega^{2}r+(\frac{r}{r_{0}})^{-j}\bigg(8(\frac{r}
{r_{0}})^{j}\xi_{2}^{2}c_4jr^{2}+8(\frac{r}{r_{0}})^{j}\xi_{2}
^{2}c_4r^{3}
\\\label{59}
&+&16(\frac{r}{r_{0}}) ^{j}\xi_{2}^{2}j+16(\frac{r}{r_{0}})^{j}
\xi_{2}^{2}r+r^{4}\bigg)^{\frac{1}{2}}\xi_{3}\bigg)\bigg\}
^{\frac{1}{2}}\bigg].
\end{eqnarray}
Figure \textbf{5} manifests that WH geometry is asymptotically flat
and shape function maintains its positivity. The right graph in the
lower panel indicates that $\frac{dh(r_{0})}{dr}<1$ and the
associated left graph shows that WH throat exist at $r_{0}=0.01$.
The last plot shows that flaring-out condition satisfied at wormhole
throat in this case. The corresponding null energy condition turns
out to be
\begin{figure}
\epsfig{file=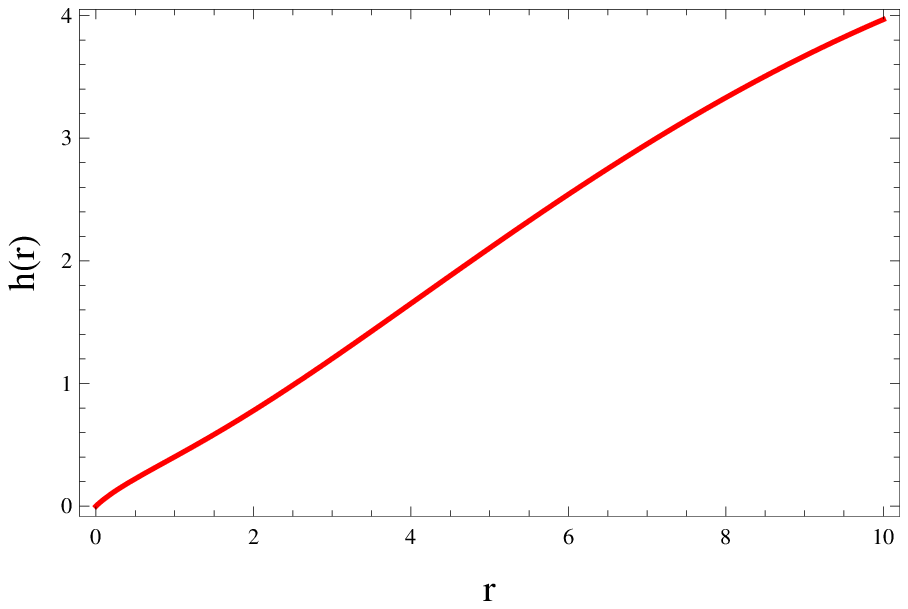,width=.5\linewidth}
\epsfig{file=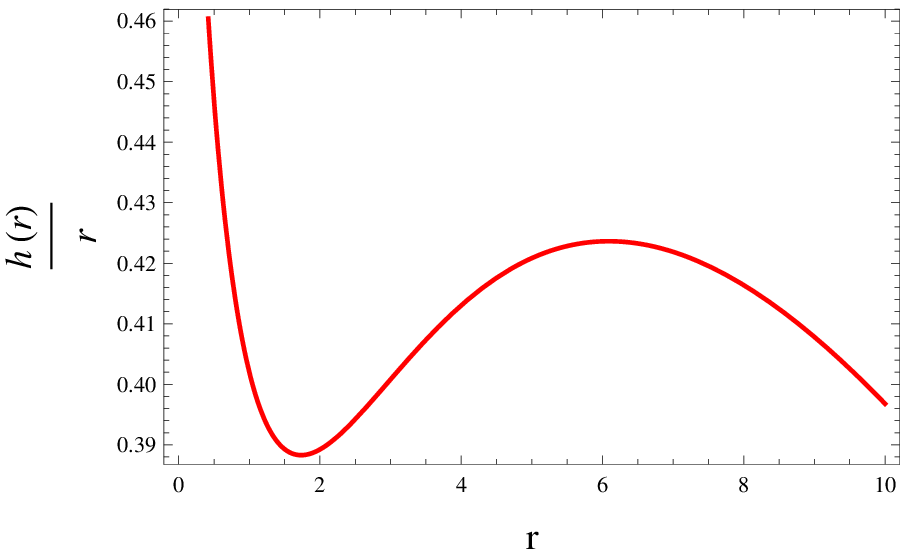,width=.53\linewidth}
\epsfig{file=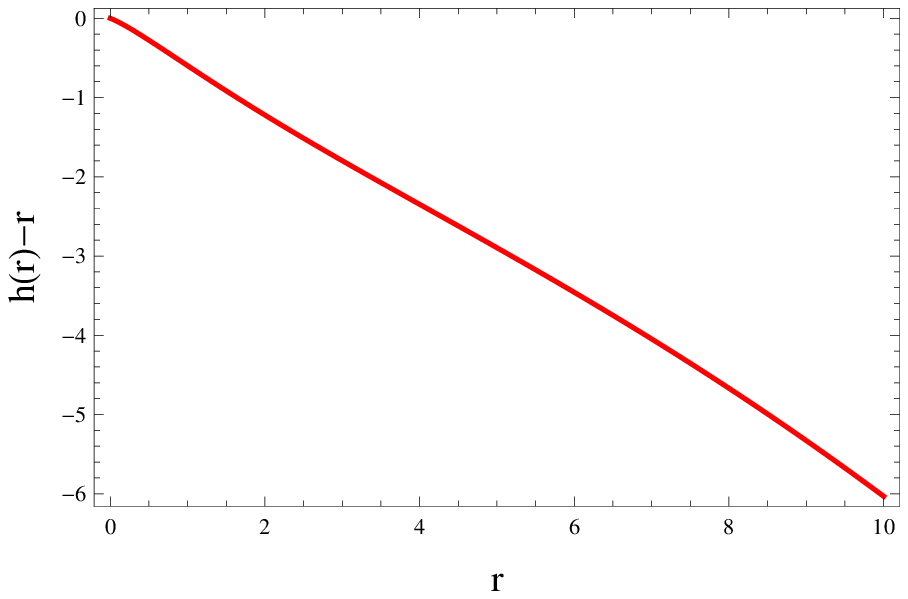,width=.5\linewidth}
\epsfig{file=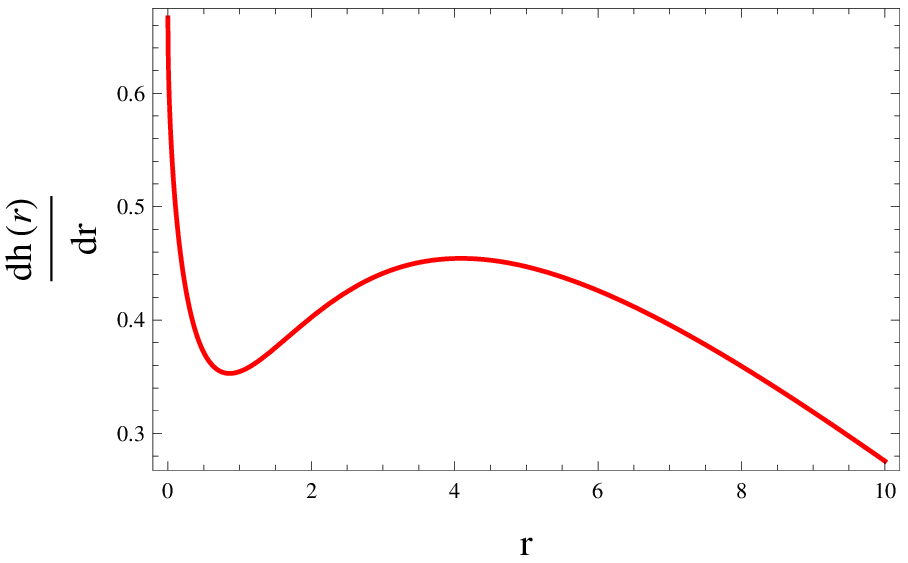,width=.53\linewidth}\center
\epsfig{file=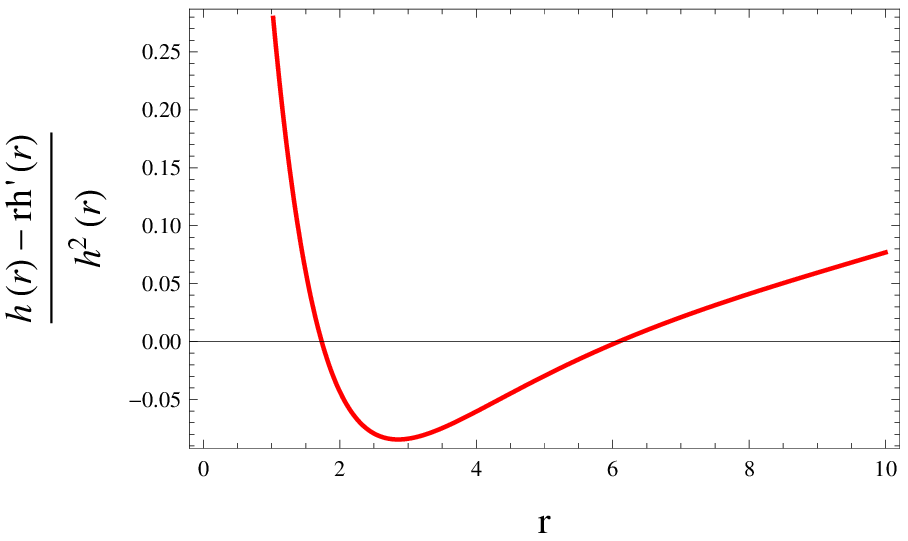,width=.53\linewidth}\caption{Graphs of WSF
versus $r$.}
\end{figure}
\begin{figure}\center
\epsfig{file=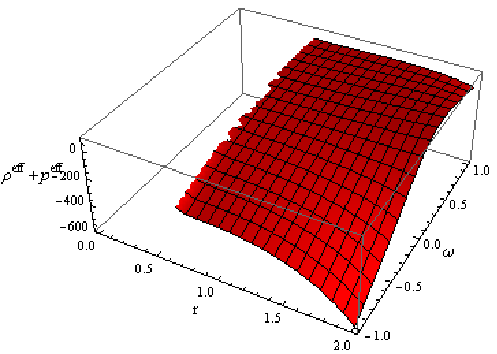,width=.5\linewidth}\caption{Graph of
$\rho^{eff}+p^{eff}$ versus $r$.}
\end{figure}
\begin{eqnarray}\nonumber
\rho^{eff}+p^{eff}&=&(1+\omega)\bigg[\frac{1}{2\xi_{2}\xi_{3}
(3\omega^{2}+1)}\bigg[-\xi_{2}\omega+\bigg\{\frac{1}{r+j}
\bigg(3\omega^{2}\xi_{3}(\frac{r}{r_{0}})^{-j}r^{2}
\\\nonumber
&+&3(\frac{r}{r_{0}})^{-j}\bigg(8(\frac{r}{r_{0}})^{j}\xi_{2}
^{2}c_4jr^{2}+8(\frac{r}{r_{0}})^{j}\xi_{2}^{2}c_4r^{3}+16
(\frac{r}{r_{0}})^{j}\xi_{2}^{2}j+16r
\\\nonumber
&\times&(\frac{r}{r_{0}})^{j}\xi_{2}^{2}+r^{4}\bigg)^{\frac{1}
{2}}\xi_{3}\omega^{2}+\xi_{3}(\frac{r}{r_{0}})^{-j}r^{2}
+\xi_{2}^{2}\omega^{2}j+\xi_{2}^{2}\omega^{2}r+(\frac{r}
{r_{0}})^{-j}
\\\nonumber
&\times&\bigg(8(\frac{r}{r_{0}})^{j}\xi_{2}^{2}c_4jr^{2}
+8(\frac{r}{r_{0}})^{j}\xi_{2}^{2}c_4r^{3}+16(\frac{r}{r_{0}})
^{j}\xi_{2}^{2}j+16(\frac{r}{r_{0}})^{j} \xi_{2}^{2}r
\\\nonumber
&+&r^{4}\bigg)^{\frac{1}{2}}\xi_{3}\bigg)\bigg\}^{\frac{1}{2}}
\bigg]\bigg]+2\xi_{3}(\omega^{2}+4\omega+1)\bigg[\frac{1}{2
\xi_{2}\xi_{3}(3\omega^{2}+1)}\bigg[-\xi_{2}\omega
\\\nonumber
&+&\bigg\{\frac{1}{r+j}\bigg(3\omega^{2}\xi_{3}(\frac{r}{r_{0}})
^{-j}r^{2}+3(\frac{r}{r_{0}})^{-j}\bigg(8(\frac{r}{r_{0}})^{j}
\xi_{2}^{2}c_4jr^{2}+8(\frac{r}{r_{0}})^{j}
\\\nonumber
&\times&\xi_{2}^{2}c_4r^{3}+16(\frac{r}{r_{0}})^{j}\xi_{2}^{2}
j+16(\frac{r}{r_{0}})^{j}\xi_{2}^{2}r+r^{4}\bigg)^{\frac{1}{2}}
\xi_{3}\omega^{2}+ \xi_{3}(\frac{r}{r_{0}})^{-j}r^{2}
\\\nonumber
&+&\xi_{2}^{2}\omega^{2}j+\xi_{2}^{2}\omega^{2}r+(\frac{r}{r_{0}
})^{-j}\bigg(8(\frac{r}{r_{0}})^{j}\xi_{2}^{2}c_4jr^{2}+8(\frac
{r}{r_{0}})^{j}\xi_{2}^{2}c_4r^{3}
\\\label{60}
&+&16(\frac{r}{r_{0}}) ^{j}\xi_{2}^{2}j+16(\frac{r}{r_{0}})^{j}
\xi_{2}^{2}r+r^{4}\bigg)^{\frac{1}{2}}\xi_{3}\bigg)\bigg\}
^{\frac{1}{2}}\bigg]\bigg]^{2}.
\end{eqnarray}
Figure \textbf{6} shows that viable traversable WH exists in the
specific range of EoS parameter.

\subsubsection*{Case II: $\alpha(r)=e^{-\frac{r_{0}}{r}}$}

Here, Eq.(\ref{59}) leads to
\begin{eqnarray}\nonumber
\beta(r)&=&2\ln\bigg[\frac{1}{4\xi_{2}}\bigg\{r^{3}+\bigg
\{8(e^{1/2(e^{\frac{r_{0}}{r}})^{-1}})^{2}\xi_{2}^{2}
\xi_{4}r^{5}+8e^{-1/2\frac{1}{r}(2r_{0}e^{\frac{r_{0}}{r}}
-r)(e^{\frac{r_{0}}{r}})^{-1}}e^{1/2(e^{\frac{r_{0}}{r}})
^{-1}}
\\\nonumber&\times&
\xi_{2}^{2}\xi_{4}r_{0}r^{3}+16(e^{1/2(e^\frac{r_{0}}{r})
^{-1}})^{2}\xi_{2}^{2}r^{3}+16e^{-1/2\frac{1}{r}(2r_{0}
e^{\frac{r_{0}}{r}}-r)(e^{\frac{r_{0}}{r}})^{-1}}e^{1/2
(e^{\frac{r_{0}}{r}})^{-1}}\xi_{2}^{2}r_{0}r
\\\label{61}&+&
r^{6}\bigg\}^\frac{1}{2}\bigg\}\bigg\{e^{1/2(e^{\frac{r_{0}
}{r}})^{-1}}r^{2}+e^{-1/2\frac{1}{r}(2r_{0}e^{\frac{r_{0}}
{r}}-r)(e^{\frac{r_{0}}{r}})^{-1}}r_{0}\bigg\}^{-1}\bigg].
\end{eqnarray}
The corresponding WSF is
\begin{eqnarray}\nonumber
h(r)&=&\bigg[1-16\xi_{2}^{2}\bigg\{e^{1/2(e^{\frac{r_{0}}{r}
})^{-1}}r^{2}+e^{-1/2\frac{1}{r}(2r_{0}e^{\frac{r_{0}}{r}}-r)
(e^{\frac{r_{0}}{r}})^{-1}}r_{0}\bigg\}^{2}\bigg[r^{3}+\bigg
\{8\xi_{2}^{2}\xi_{4}r^{5}
\\\nonumber
&\times&(e^{1/2(e^{\frac{r_{0}}{r}})^{-1}})^{2}+8e^{-1/2{
\frac{1}{r}(2r_{0}e^{\frac{r_{0}}{r}}-r)(e^{\frac{r_{0}}{r}}
)^{-1}}}e^{1/2(e^{\frac{r_{0}}{r}})^{-1}}\xi_{2}^{2}\xi_{4}
r_{0}r^{3}+16\xi_{2}^{2}r^{3}
\\\nonumber
&\times&(e^{1/2(e^{\frac{r_{0}}{r}})^{-1}})^{2}+16e^{-1/2\frac
{1}{r}(2r_{0}e^{\frac{r_{0}}{r}}-r)(e^{\frac{r_{0}}{r}})^{-1}}
e^{1/2(e^{\frac{r_{0}}{r}})^{-1}}\xi_{2}^{2}r_{0}r
\\\label{62}
&+&r^{6}\bigg\}^{\frac{1}{2}}\bigg] ^{-2}\bigg]r.
\end{eqnarray}
Figure \textbf{7} determines that WSF is positive with $h(r)<r$ but
the behavior of wormhole is not asymptotically flat. The WH throat
is located at $r_{0} = 0.001$ with $\frac {dh(r_{0})}{dr}<1$. The
flaring-out condition is fulfilled at wormhole throat. The energy
density turns out to be
\begin{figure}
\epsfig{file=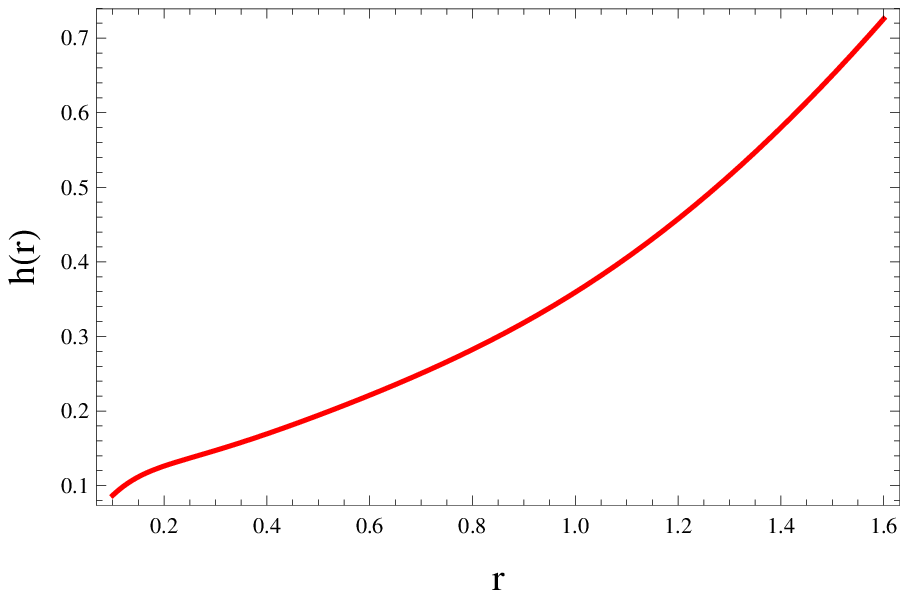,width=.5\linewidth}
\epsfig{file=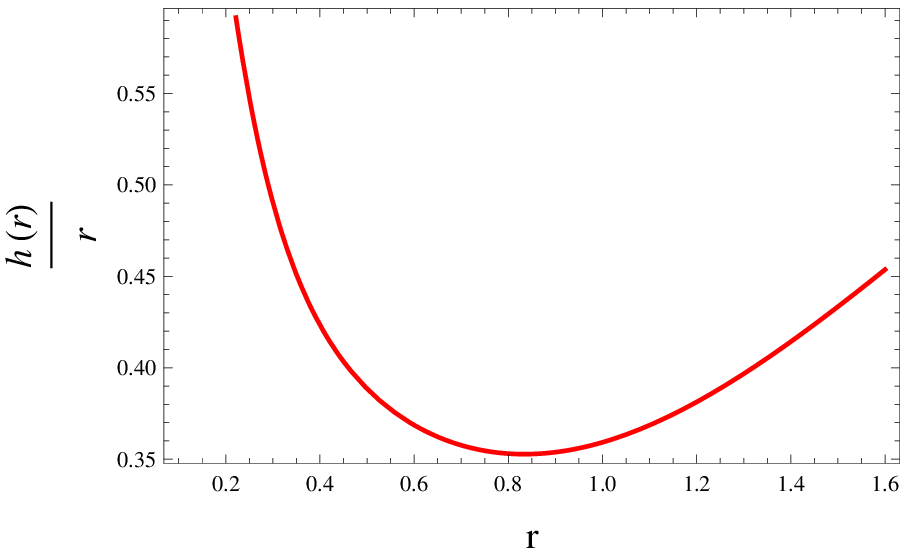,width=.53\linewidth}
\epsfig{file=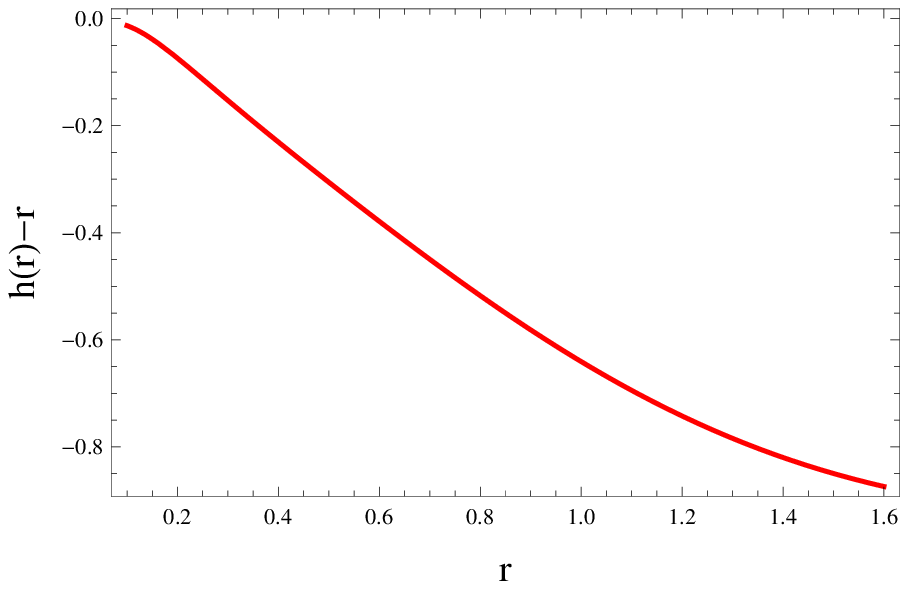,width=.5\linewidth}
\epsfig{file=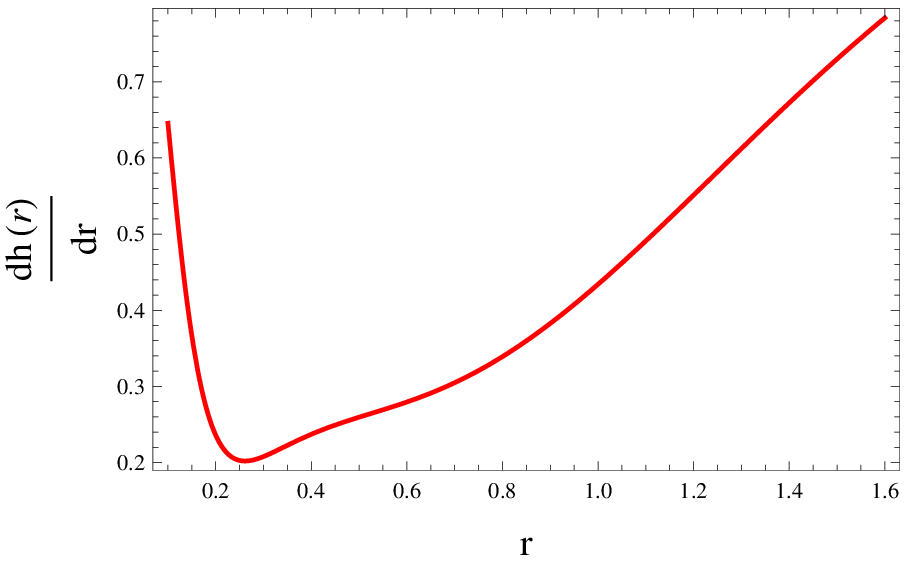,width=.53\linewidth}\center
\epsfig{file=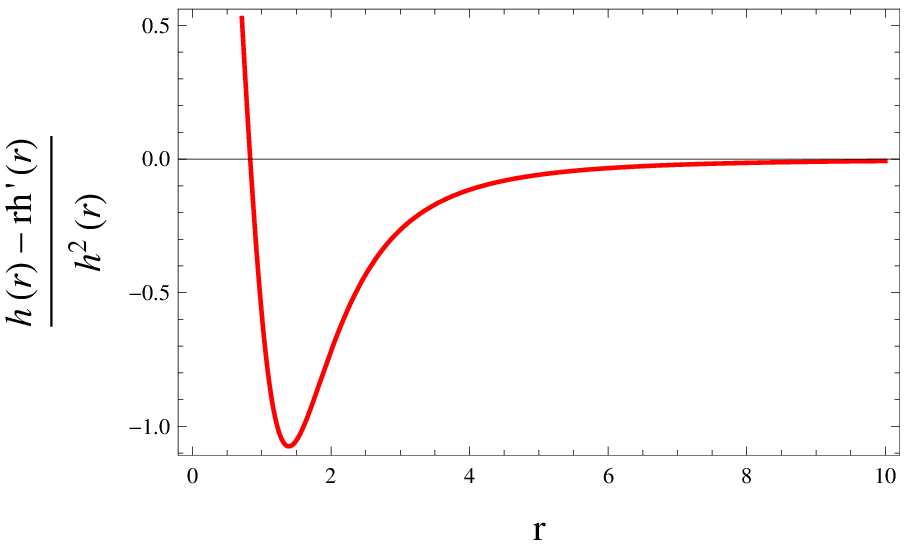,width=.53\linewidth}\caption{Graphs of WSF
versus $r$.}
\end{figure}
\begin{eqnarray}\nonumber
\rho&=&\frac{1}{2\xi_{2}\xi_{3}(3\omega^{2}+1)}\bigg[-\xi_{2}
\omega+\bigg[\xi_{2}^{2}\bigg\{48\xi_{3}\omega^{2}r^{2}+48
\xi_{3}r_{0}\omega^{2}e^{-\frac{r_{0}}{r}}+16\xi_{3}r^{2}
\\\nonumber
&+&\omega^{2}r^{3}+16\xi_{3}r_{0}e^{-\frac{r_{0}}{r}}+\bigg
\{r\bigg(8e^{e^{-\frac{r_{0}}{r}}}\xi_{2}^{2}\xi_{4}r^{4}+8
e^{-\frac{1}{r}e^{-\frac{r_{0}}{r}}(r_{0}e^{\frac{r_{0}}{r}}
-r)}\xi_{2}^{2}\xi_{4}r_{0}r^{2}
\\\nonumber
&+&16e^{e^{-\frac{r_{0}}{r}}}\xi_{2}^{2}r^{2}+r^{5}+16e^{-\frac
{1}{r}e^{-\frac{r_{0}}{r}}(r_{0}e^{\frac{r_{0}}{r}}-r)}\xi_{2}
^{2}r_{0}\bigg)\bigg\}^{\frac{1}{2}}\omega^{2}\bigg\}\bigg\{r^
{3}+\bigg\{r\bigg(8r^{4}
\\\nonumber
&\times& e^{e^{-\frac{r_{0}}{r}}}\xi_{2}^{2}\xi_{4}+8\xi_{2}^{2}
\xi_{4}r_{0}r^{2}e^{-\frac{1}{r}e^{-\frac{r_{0}}{r}}(r_{0}
e^{\frac{r_{0}}{r}}-r)}+16e^{e^{-\frac{r_{0}}{r}}}\xi_{2}^{2}
r^{2}+r^{5}+16\xi_{2}^{2}r_{0}
\\\label{63}
&\times&
e^{-\frac{1}{r}e^{-\frac{r_{0}}{r}}(r_{0}e^{\frac{r_{0}}{r}}-r)}
\bigg)\bigg\}^{\frac{1}{2}}\bigg\}^{-1}\bigg]^{\frac{1}{2}}\bigg].
\end{eqnarray}
The null energy condition yields
\begin{eqnarray}\nonumber
\rho^{eff}+p^{eff}&=&(1+\omega)\Bigg[\frac{1}{2\xi_{2}\xi_{3}(3\omega
^{2}+1)}\bigg[-\xi_{2}\omega+\bigg[\xi_{2}^{2}\bigg\{48\xi_{3}\omega
^{2}r^{2}+48\xi_{3}r_{0}\omega^{2}e^{-\frac{r_{0}}{r}}
\\\nonumber
&+&16\xi_{3}r^{2}+\omega^{2}r^{3}+16\xi_{3}r_{0}e^{-\frac{r_{0}}{r}}
+\bigg\{r\bigg(8e^{e^{-\frac{r_{0}}{r}}}\xi_{2}^{2}\xi_{4}r^{4}+8
e^{-\frac{1}{r}e^{-\frac{r_{0}}{r}}(r_{0}e^{\frac{r_{0}}{r}}-r)}
\\\nonumber
&\times&\xi_{2}^{2}\xi_{4}r_{0}r^{2}+16e^{e^{-\frac{r_{0}}{r}}}
\xi_{2}^{2}r^{2}+r^{5}+16e^{-\frac{1}{r}e^{-\frac{r_{0}}{r}}(r_{0}
e^{\frac{r_{0}}{r}}-r)}\xi_{2}^{2}r_{0}\bigg)\bigg\}^{\frac{1}{2}}
\omega^{2}\bigg\}\bigg\{r^{3}
\\\nonumber
&+&\bigg\{r\bigg(8e^{e^{-\frac{r_{0}}{r}}}\xi_{2}^{2}\xi_{4}r^{4}
+8\xi_{2}^{2}\xi_{4}r_{0}r^{2}e^{-\frac{1}{r}e^{-\frac{r_{0}}{r}}
(r_{0}e^{\frac{r_{0}}{r}}-r)}+16e^{e^{-\frac{r_{0}}{r}}}\xi_{2}^{2}
r^{2}+r^{5}
\\\nonumber
&+&16e^{-\frac{1}{r}{e^{-\frac{r_{0}}{r}}}(r_{0}e^{\frac{r_{0}}{r}}
-r)}\xi_{2}^{2}r_{0}\bigg)\bigg\}^{\frac{1}{2}}\bigg\}^{-1}\bigg]
^{\frac{1}{2}}\bigg]\Bigg]+2\xi_{3}(3\omega^{2}+4\omega+1)
\\\nonumber
&\times& \Bigg[\frac{1}{2\xi_{2}\xi_{3}(3\omega^{2}+1)}\bigg[-
\xi_{2}\omega+\bigg[\xi_{2}^{2}\bigg\{48\xi_{3}\omega^{2}r^{2}+48
\xi_{3}r_{0}\omega^{2}e^{-\frac{r_{0}}{r}}+16\xi_{3}r^{2}
\\\nonumber
&+&\omega^{2}r^{3}+16\xi_{3}r_{0}e^{-\frac{r_{0}}{r}}+\bigg\{r\bigg
(8e^{e^{-\frac{r_{0}}{r}}}\xi_{2}^{2}\xi_{4}r^{4}+8e^{-\frac{1}{r}
e^{-\frac{r_{0}}{r}}(r_{0}e^{\frac{r_{0}}{r}}-r)}\xi_{2}^{2}\xi_{4}
r_{0}r^{2}
\\\nonumber
&+&16e^{e^{-\frac{r_{0}}{r}}}\xi_{2}^{2}r^{2}+r^{5}+16e^{-\frac{1}{r}
e^{-\frac{r_{0}}{r}}(r_{0}e^{\frac{r_{0}}{r}}-r)}\xi_{2}^{2}r_{0}\bigg)
\bigg\}^{\frac{1}{2}}\omega^{2}\bigg\}\bigg\{r^{3}+\bigg\{r\bigg(8r^{4}
\\\nonumber
&\times&e^{e^{-\frac{r_{0}}{r}}}\xi_{2}^{2}\xi_{4}+8\xi_{2}^{2}\xi_{4}
r_{0}r^{2}e^{-\frac{1}{r}e^{-\frac{r_{0}}{r}}(r_{0}e^{\frac{r_{0}}{r}}
-r)}+16e^{e^{-\frac{r_{0}}{r}}}\xi_{2}^{2}r^{2}+r^{5}+16\xi_{2}^{2}r_{0}
\\\label{64}
&\times&e^{-\frac{1}{r}e^{-\frac{r_{0}}{r}}(r_{0}e^{\frac{r_{0}}{r}}-r)}
\bigg)\bigg\}^{\frac{1}{2}}\bigg\}^{-1}\bigg]^{\frac{1}{2}}\bigg]\Bigg]^{2}.
\end{eqnarray}
The graphical behavior of effective matter variables is given in
Figure \textbf{8} which shows that traversable WH exists in this
modified theory for $-1\leq\omega\leq0$.
\begin{figure}\center
\epsfig{file=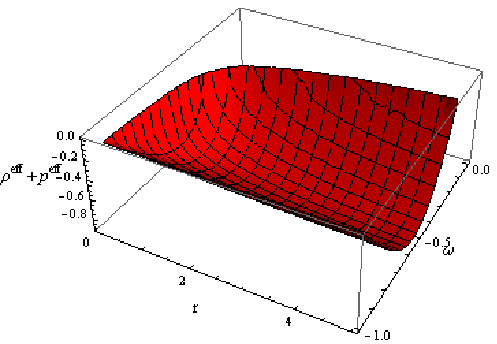,width=.5\linewidth}\caption{Graph of
$\rho^{eff}+p^{eff}$ versus $r$.}
\end{figure}

\section{Stability Analysis}

In order to analyze the valid and consistent cosmic structures,
stability is significant. It is more interesting to examine cosmic
objects that display stable behavior under the external
perturbations. In the following, we investigate the stable WH
solutions via \emph{causality condition} and \emph{adiabatic index}.

\subsection{Causality Condition}

Stable stellar system is considered the more viable in the realm of
gravitational physics. When the system experiences non-disappearing
forces, it is important to observe how the matter variables behave
after disruption from the equilibrium condition. The causality
condition is a mathematical requirement that imposes constraints on
the system. In the context of stability analysis, the causality
condition is used to check whether a system is stable or not. If the
system satisfies the causality condition, it means that the
corresponding system is stable and that the output will not exhibit
any oscillations or instability. Here, we use the causality
condition to check the stable state of WH solutions. According to
this condition, the square speed of sound
$(u_{s}^{2}=\frac{dp^{eff}}{d\rho^{eff}})$ should satisfy the limit,
$0\leq u_{s}^{2}<1$ \cite{41a}.

However, outside this region of stability, the output of the system
may exhibit different forms of instability. For example, if the
function violates the causality condition, it may give rise to an
unstable system, and that exhibits oscillations or even divergent
behavior. In some cases, the instability may be in the form of noise
or other unwanted behavior that can make the system unusable for its
intended purpose. Another form of instability that can occur outside
the region of stability is related to the Nyquist stability
criterion. This criterion is based on the mapping of the frequency
response of the function onto the complex plane. If the Nyquist plot
encircles the point (-1,0) in a clockwise direction, the system is
unstable. This instability can manifest itself as oscillations or
other types of unwanted behavior. Hence, while the causality
condition is an important requirement for stability, it is not the
only factor that determines whether a system is stable or not. Other
forms of instability can occur outside the region of stability, and
these may require different methods for analysis and control.
Figures \textbf{9} and \textbf{10} show that WH solutions satisfy
the required causality condition in the presence of modified terms.
\begin{figure}
\epsfig{file=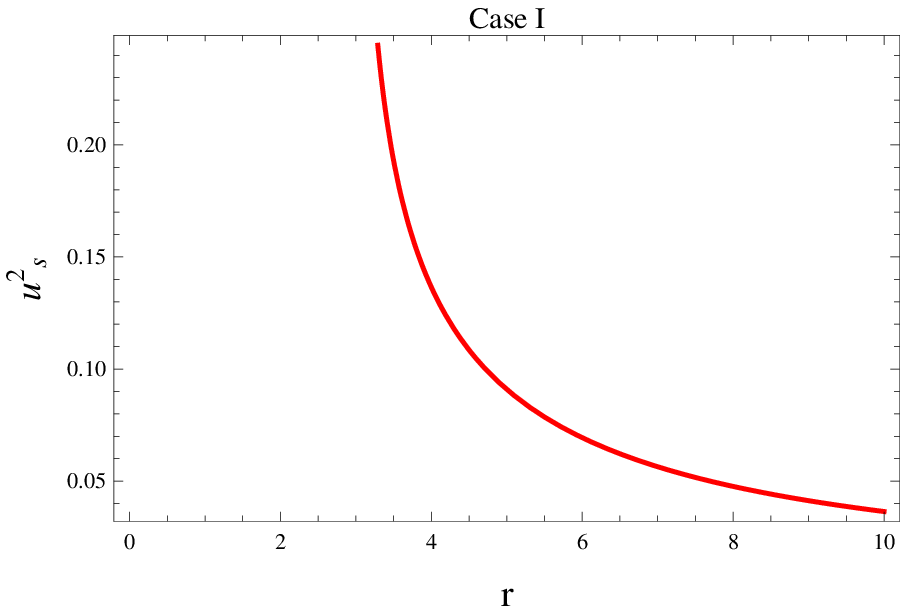,width=.5\linewidth}
\epsfig{file=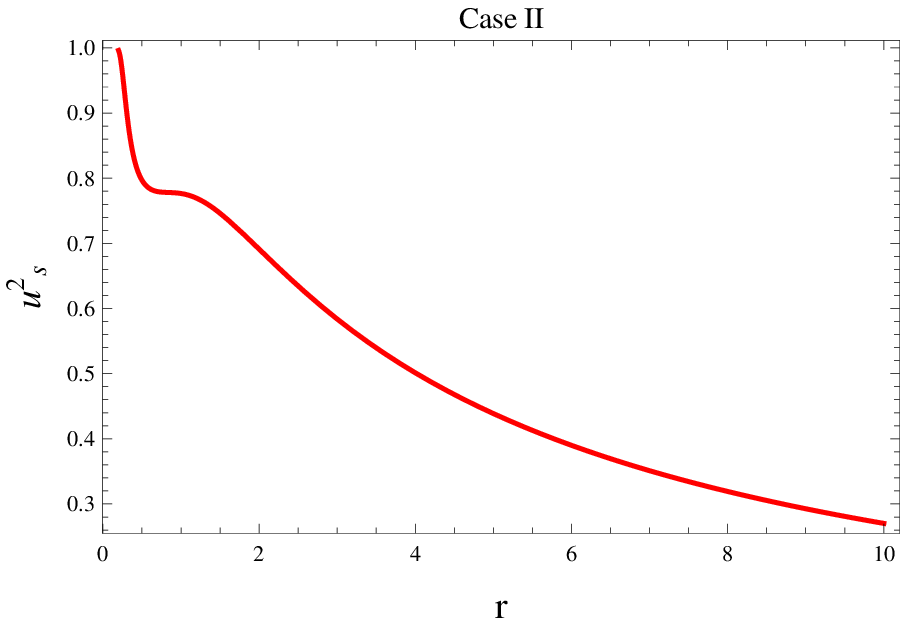,width=.5\linewidth}\caption{Graphs of sound
speed versus $r$ for dust case.}
\end{figure}
\begin{figure}
\epsfig{file=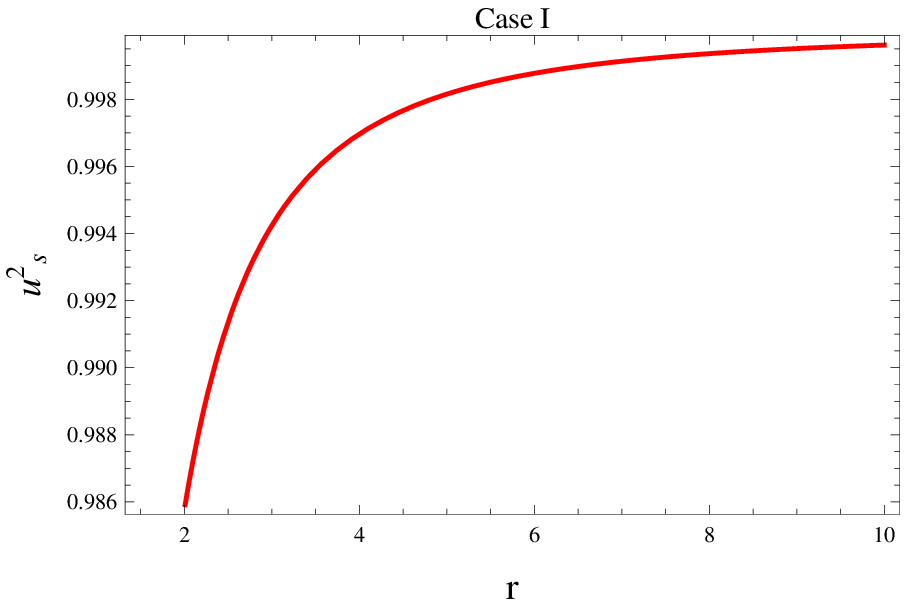,width=.5\linewidth}
\epsfig{file=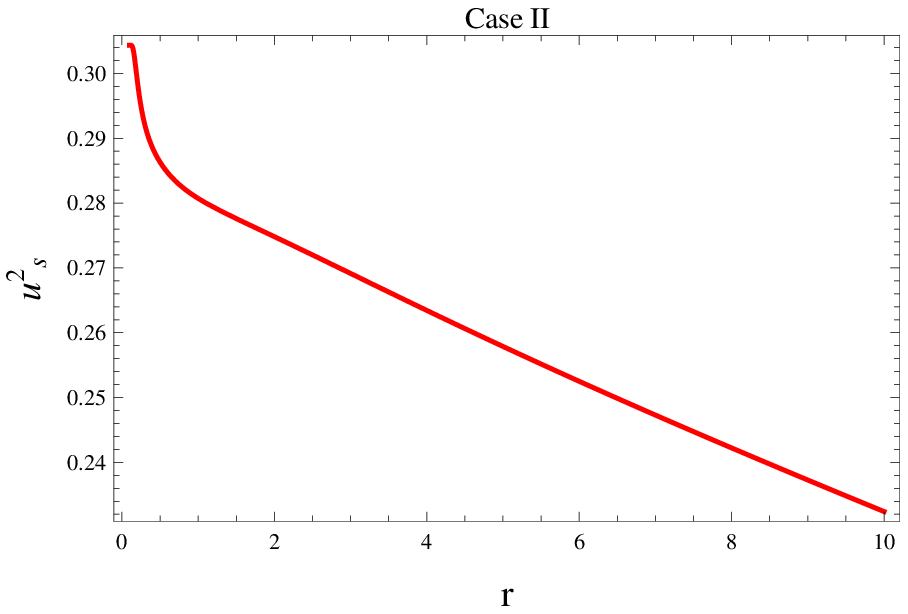,width=.5\linewidth}\caption{Graphs of sound
speed versus $r$ for non dust case.}
\end{figure}

\subsection{Adiabatic Index}

This is an alternative technique to explore the stability of
celestial objects. The adiabatic index, also known as the gamma
factor, is a measure of the thermodynamic properties of a gas and is
defined as the ratio of the specific heat at constant pressure to
the specific heat at constant volume. In the context of
astrophysics, the adiabatic index is used to determine the stability
of a star against radial perturbation. A star is said to be stable
if it can resist small radial oscillations without undergoing
collapse or explosion. The adiabatic index is related to the speed
of sound in the stellar material, and a lower value of the adiabatic
index indicates a softer material and a higher value indicates a
stiffer material. For a stable star, the adiabatic index must be
greater than a critical value, typically around 4/3 \cite{42}. The
adiabatic index depends on the composition of the stellar material,
which in turn affects the nuclear reactions that take place in the
star's core. As a star burns through its fuel, the composition of
its material changes, and this can affect the star's adiabatic index
and stability. Therefore, the adiabatic index is an important
parameter in the study of stellar structure and evolution, and it is
used to understand the behavior of stars and their compositions.
Many researchers used this condition in the literature \cite{43}.

The adiabatic index is expressed as
\begin{equation}\label{65}
\Gamma^{eff}=\frac{p^{eff}+\rho^{eff}} {p^{eff}}u_{s}^{2}.
\end{equation}
According to Heintzmann and Hillebrandt \cite{42}, a system is
stable if $\Gamma>4/3$, otherwise it is unstable. Figures
\textbf{11} and \textbf{12} show that WH solutions satisfy the
required limits in both (dust and non-dust) cases, indicating that
our system is in a stable state.
\begin{figure}
\epsfig{file=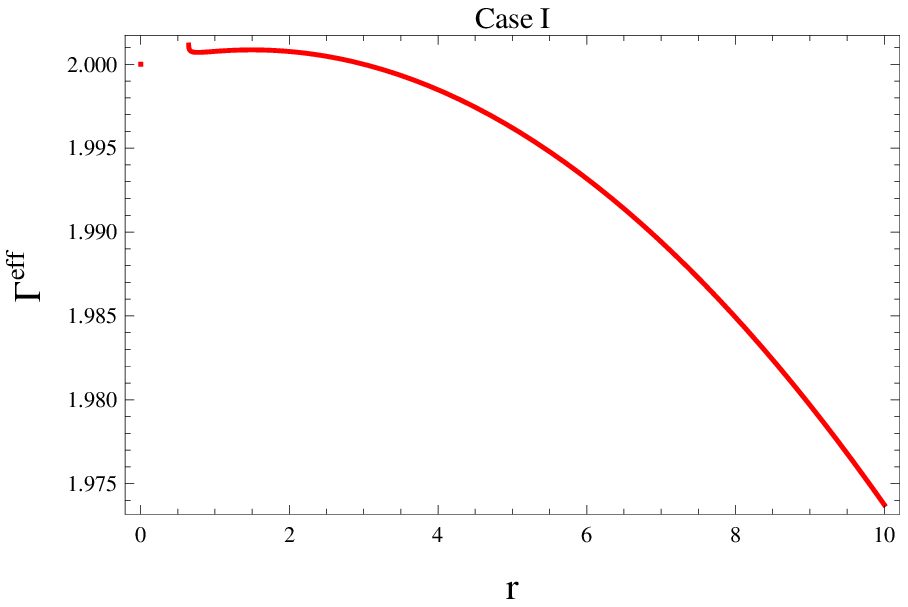,width=.5\linewidth}
\epsfig{file=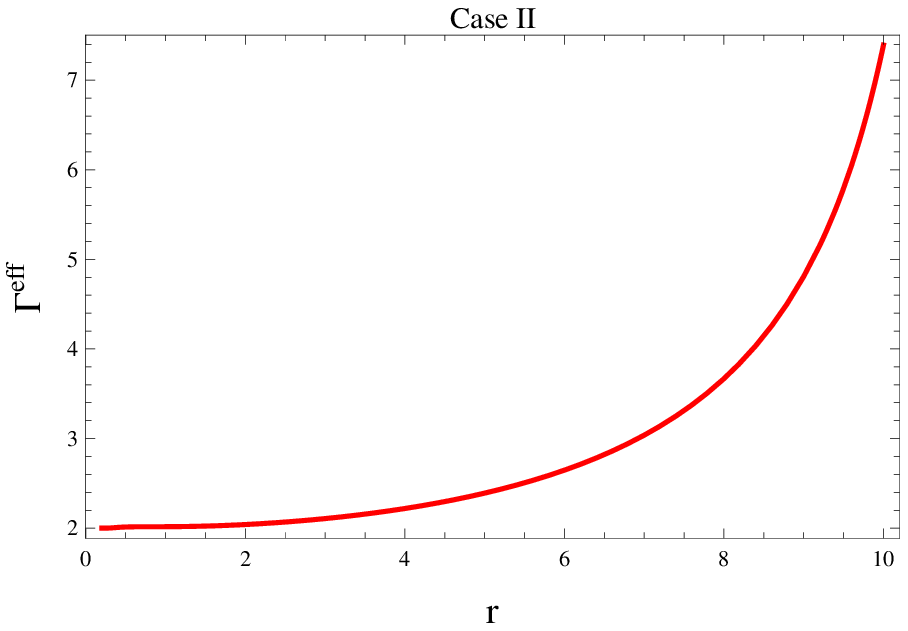,width=.5\linewidth}\caption{Graphs of adiabatic
index versus $r$ for dust case.}
\end{figure}
\begin{figure}
\epsfig{file=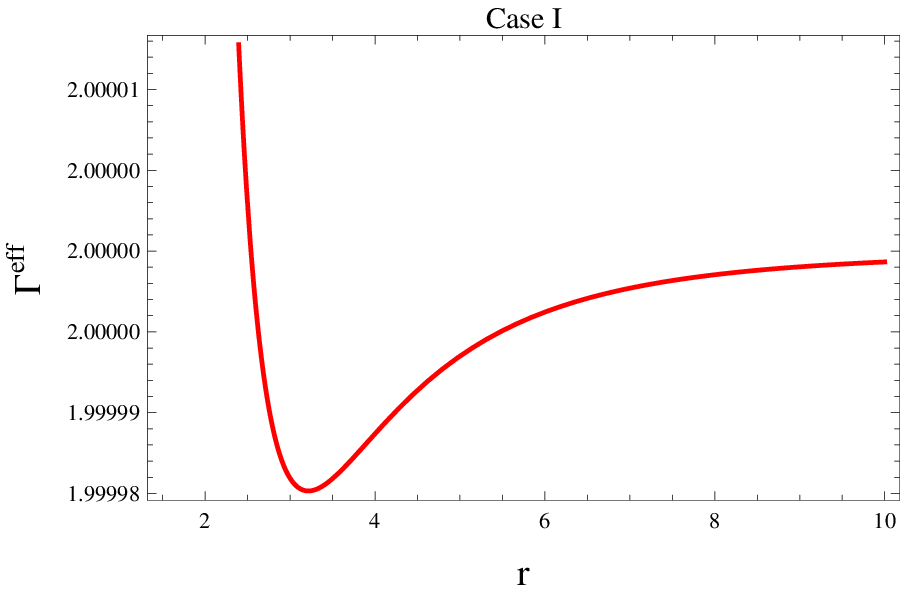,width=.5\linewidth}
\epsfig{file=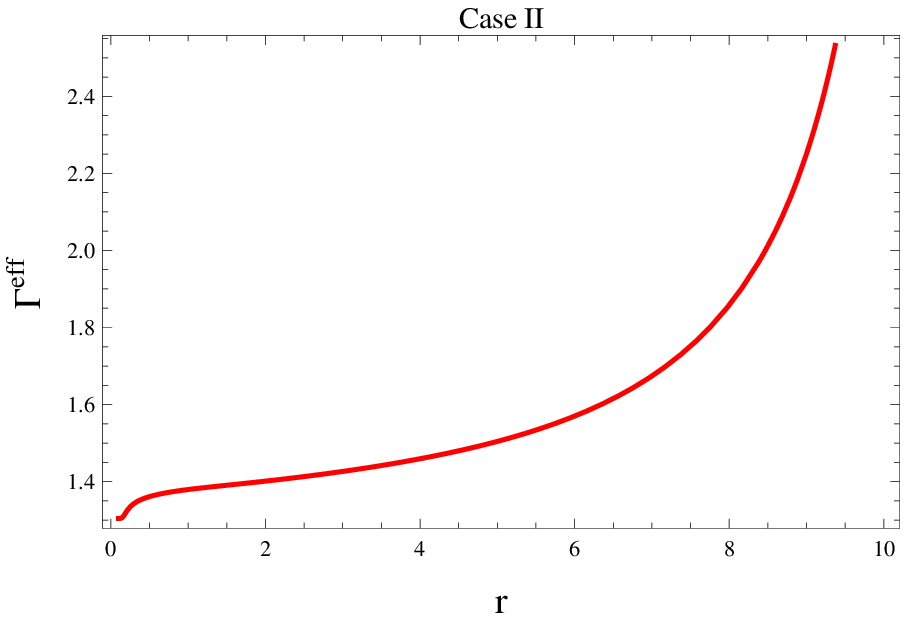,width=.5\linewidth}\caption{Plots of adiabatic
index versus $r$ for non dust case.}
\end{figure}

\section{Final Remarks}

The existence of WH structure is a crucial issue in the field of
astrophysics. In GR, the existence of exotic matter is significant
for the presence of physically realistic WH geometry. In the last
few decades, the scientific community has paid a lot of attention to
modified theories of gravity as a possible alternative to GR. Many
scientists found these modified theories interesting to analyze the
viable traversable WH geometry due to the violation of energy bounds
by the effective stress-energy tensor which ensures the presence of
viable WH structure.

In this manuscript, we have used the NS approach to find some exact
solutions that help to formulate static WH solutions in
$f(\mathcal{R},\mathcal{T}^2)$ theory. We have examined the
existence of exotic matter in WHs via violation of null energy
condition. For different matter configurations, we have investigated
the viable WH geometry corresponding to different redshift
functions, i.e., $\alpha(r)=j\ln(\frac{r}{r_{0}})$ and
$\alpha(r)=e^{-\frac{r_{0}}{r}}$. Finally, we have investigated the
stability of the obtained WH solutions through causality condition
and adiabatic index. We have examined NS generators and conserved
quantities corresponding to both dust and non-dust cases. The
summary of the obtained results is given as follows.
\begin{itemize}
\item
For $\alpha(r)=j\ln(\frac{r}{r_{0}})$, it is found that WSF
satisfies all the required conditions and preserves asymptotically
flat behavior for both dust and dust matter configurations (Figures
\textbf{1} and \textbf{5}).
\item
Wormhole shape function does not preserve asymptotically flat
behavior for $\alpha(r)=e^{-\frac{r_{0}}{r}}$ (Figures \textbf{3}
and \textbf{7}).
\item
For dust matter configuration, the effective energy-momentum tensor
violates the null energy condition for both choices of redshift
functions that show the existence of traversable WH geometry in
$f(\mathcal{R},\mathcal{T}^2)$ theory (Figures \textbf{2}and
\textbf{4}).
\item
In the non-dust case, we have found that $\rho^{eff}+p^{eff}\leq0$
for a specific range of EoS parameter which indicates the existence
of viable and traversable WH (Figures \textbf{6} and \textbf{8}).
\item
It is found that traversable WHs are stable for both types of
redshift function in the presence of modified terms (Figures
\textbf{9}- \textbf{12}).
\end{itemize}

In the framework of GR, Fayyaz and Shamir \cite{44} studied
physically realistic and stable WH solutions in the presence of
exotic matter. The same authors \cite{45} found that the considered
WSF satisfies the null energy condition and hence shows the absence
of exotic matter which yields non-traversable WH geometry in
$f(\mathcal{R})$ theory. We have found that the null energy
condition is violated in the context of
$f(\mathcal{R},\mathcal{T}^{2})$ theory, which shows that viable and
stable traversable WH solutions exist in this modified theory.

\vspace{0.25cm}

\section*{Appendix A}
\renewcommand{\theequation}{A\arabic{equation}}
\setcounter{equation}{0}

The coefficients of Eq.(\ref{22}) are
\begin{eqnarray}\label{25}
&&\Psi_{,\beta}=0, \quad \lambda_{,\alpha}=0, \quad
\lambda_{,\beta}=0, \quad \lambda_{,\eta}=0, \quad
\lambda_{,\mathcal{R}}=0, \quad \lambda_{,\mathcal{T}
^{2}}=0,\\\label{26}
&&\eta\gamma^{1}_{,\beta}f_{\mathcal{R}\mathcal{R}}
+2\gamma^{3}_{,\beta}f_{\mathcal{R}\mathcal{R}}=0,
\\\label{27}
&&\eta\gamma^{1}_{,\mathcal{R}}f_{\mathcal{R}\mathcal
{R}}+2\gamma^{3}_{,\mathcal{R}}f_{\mathcal{R}\mathcal{R}}=0,
\\\label{28}
&&\eta\gamma^{1}_{,\beta}f_{\mathcal{R}\mathcal{T}^{2}}
+2\gamma^{3}_{,\beta}f_{\mathcal{R}\mathcal{T}^{2}}=0,
\\\label{29}
&&\eta\gamma^{1}_{,\mathcal{T}^{2}}f_{\mathcal{R}\mathcal
{T}^{2}}+2\gamma^{3}_{,\mathcal{T}^{2}}f_{\mathcal{R}
\mathcal{T}^{2}}=0,
\\\label{30}
&&\gamma^{3}_{,\beta}f_{\mathcal{R}}+\eta\gamma^{4}_{,\beta}
f_{\mathcal{R}\mathcal{R}}+\eta\gamma^{5}_{,\beta}f_{\mathcal
{R}\mathcal{T}^{2}}=0,
\\\label{31}
&&\eta\gamma^{1}_{,r}f_{\mathcal{R}\mathcal{R}}+2\gamma^{3}
_{,r}f_{\mathcal{R}\mathcal{R}}-e^{\frac{\beta-\alpha}{2}}
\Psi_{,\mathcal{R}}=0,
\\\label{32}
&&\gamma^{3}_{,\alpha}f_{\mathcal{R}}+\eta\gamma^{4}_{,\alpha}
f_{\mathcal{R}\mathcal{R}}+\eta\gamma^{5}_{,\alpha}f_{\mathcal
{R}\mathcal{T}^{2}}=0,
\\\label{33}
&&\eta\gamma^{1}_{,r}f_{\mathcal{R}\mathcal{T}^{2}}+2\gamma^{3}
_{,r}f_{\mathcal{R}\mathcal{T}^{2}}-e^{\frac{\beta-\alpha}{2}}
\Psi_{,\mathcal{T}^{2}}=0,
\\\label{34}
&&\gamma^{1}_{,\beta}f_{\mathcal{R}}+\gamma^{3}_{,\beta}\eta^{-1}
f_{\mathcal{R}}+2\gamma^{4}_{,\beta}f_{\mathcal{R}\mathcal{R}}
+2\gamma^{5}_{,\beta}f_{\mathcal{R}\mathcal{T}^{2}}=0,
\\\label{35}
&&\gamma^{3}_{,r}f_{\mathcal{R}}+\eta\gamma^{5}_{,r}f_{\mathcal
{R}\mathcal{R}}+\eta\gamma^{5}_{,r}f_{\mathcal{R}\mathcal{T}^{2}}
-e^{\frac{\beta-\alpha}{2}}\Psi_{,\alpha}=0,
\\\label{36}
&&\eta\gamma^{1}_{,\mathcal{T}^{2}}f_{\mathcal{R}\mathcal{R}}
+2\gamma^{3}_{,\mathcal{T}^{2}}f_{\mathcal{R}\mathcal{R}}+\eta
\gamma^{1}_{,\mathcal{R}}f_{\mathcal{R}\mathcal{T}^{2}}+2\gamma
^{3}_{,\mathcal{R}}f_{\mathcal{R}\mathcal{T}^{2}}=0,
\\\label{37}
&&\gamma^{1}_{,r}f_{\mathcal{R}}+\gamma^{3}_{,r}\eta^{-1}
f_{\mathcal{R}}+2\gamma^{4}_{,r}f_{\mathcal{R}\mathcal{R}}
+2\gamma^{5}_{,r}f_{\mathcal{R}\mathcal{T}^{2}}-e^{\frac{\beta
-\alpha}{2}}\Psi_{,\eta}=0,
\\\nonumber
&&(\gamma^{1}-\gamma^{2}-2\eta^{-1}\gamma^{3}+4\eta\gamma^{1}
_{,\eta}+4\gamma^{3}_{,\eta} -2\lambda_{,r})f_{\mathcal{R}}
\\\label{38}
&&+(2\gamma^{4}+8\eta\gamma^{4}_{,\eta})f_{\mathcal{R}\mathcal{R}}
+(2\gamma^{5}+8\eta\gamma^{5}_{,\eta})f_{\mathcal{R}\mathcal{T}^{2}}=0,
\\\nonumber
&&(2\gamma^{4}+2\eta\gamma^{4}_{,\eta}+4\gamma^{4}_{,\alpha})
f_{\mathcal{R}\mathcal{R}}+(2\gamma^{5}+2\eta\gamma^{5}_{,\eta}
+4\gamma^{5}_{,\alpha})f_{\mathcal{R}\mathcal{T}^{2}}
\\\label{39}
&&+(\gamma^{1}-\gamma^{2}+2\gamma^{1}_{,\alpha}+2\eta^{-1}\gamma
^{3}_{,\alpha}+2\gamma^{3}_{,\eta}-2\lambda_{,r})f_{\mathcal{R}}=0,
\\\nonumber
&&(\gamma^{1}-\gamma^{2}+\eta\gamma^{1}_{,\eta}+2\gamma^{3}_{,\eta}
+2\gamma^{4}_{,\mathcal{R}}-2\lambda_{,r})f_{\mathcal{R}\mathcal{R}}
+2\gamma^{4}f_{\mathcal{R}\mathcal{R}\mathcal{R}}
\\\label{40}
&&+(\gamma^{1}_{,\mathcal{R}}+\eta^{-1}\gamma^{3}_{,\mathcal{R}})
f_{\mathcal{R}}+2\gamma^{5}f_{\mathcal{R}\mathcal{R}\mathcal{T}^{2}}
+2\gamma^{5}_{,\mathcal{R}}f_{\mathcal{R}\mathcal{T}^{2}}=0,
\\\nonumber
&&(\gamma^{1}-\gamma^{2}+\eta\gamma^{1}_{,\eta}+2\gamma^{3}_{,\eta}
+2\gamma^{5}_{,\mathcal{T}^{2}}-2\lambda_{,r})f_{\mathcal{R}\mathcal
{T}^{2}}+2\gamma^{4}f_{\mathcal{R}\mathcal{R}\mathcal{T}^{2}}
\\\label{41}
&&+(\gamma^{1}_{,\mathcal{T}^{2}}+\eta^{-1}\gamma^{3}_{,\mathcal{T}
^{2}})f_{\mathcal{R}}+2\gamma^{5}f_{\mathcal{R}\mathcal{T}^{2}
\mathcal{T}^{2}}+2\gamma^{4}_{,\mathcal{T}^{2}}f_{\mathcal{R}
\mathcal{R}}=0,
\\\nonumber
&&(\eta\gamma^{1}-\eta\gamma^{2}+2\gamma^{3}+2\eta\gamma^{1}_{,\alpha}
+4\gamma^{3}_{,\alpha}+2\eta\gamma^{4}_{,\mathcal{R}}-2\eta\lambda
_{,r})f_{\mathcal{R}\mathcal{R}}
\\\label{42}
&&+2\gamma^{3}_{,\mathcal{R}}f_{\mathcal{R}}+2\eta\gamma^{4}
f_{\mathcal{R}\mathcal{R}\mathcal{R}}+2\eta\gamma^{5}
f_{\mathcal{R}\mathcal{R}\mathcal{T}^{2}}+2\gamma^{5}_{,\mathcal{R}}
f_{\mathcal{R}\mathcal{T}^{2}}=0,
\\\nonumber
&&(\eta\gamma^{1}-\eta\gamma^{2}+2\gamma^{3}+2\eta\gamma^{1}_{,\alpha}
+4\gamma^{3}_{,\alpha}+2\eta\gamma^{4}_{,\mathcal{R}}-2\eta
\lambda_{,r})f_{\mathcal{R}\mathcal{T}^{2}}
\\\label{43}
&&+2\gamma^{3}_{,\mathcal{T}^{2}}f_{\mathcal{R}}+2\eta\gamma^{4}
f_{\mathcal{R}\mathcal{R}\mathcal{T}^{2}}+2\eta\gamma^{5}
f_{\mathcal{R}\mathcal{T}^{2}\mathcal{T}^{2}}+2\gamma^{5}
_{,\mathcal{T}^{2}}f_{\mathcal{R}\mathcal{R}}=0,
\\\nonumber
&&e^{\frac{\alpha+\beta}{2}}\eta\bigg[(f-\mathcal{R}
f_{\mathcal{R}}+p+f_{\mathcal{T}^{2}}(3p^{2}+\rho^{2}
-\mathcal{T}^{2})+2\eta^{-1}f_{\mathcal{R}})
\\\nonumber
&&\times(\frac{\gamma^{1}+\gamma^{2}}{2}+\lambda_{,r})+\gamma
^{1}(f_{\mathcal{T}^{2}}(6pp_{,\alpha}+2\rho\rho_{,\alpha})
+p_{,\alpha})
\\\nonumber
&&+\gamma^{2}(f_{\mathcal{T}^{2}}(6pp_{,\beta}+2\rho\rho
_{,\beta})+p_{,\beta})+\gamma^{3}(f_{\mathcal{T}^{2}}
(6pp_{,\eta}+2\rho
\\\nonumber
&&\times\rho_{,\eta})+p_{,\eta})+\frac{\gamma^{3}}{\eta}
(f-\mathcal{R}f_{\mathcal{R}}+p+f_{\mathcal{T}^{2}}(3(p)^{2}
+(\rho)^{2}-\mathcal{T}^{2}))
\\\nonumber
&&-\gamma^{4}(f_{\mathcal{R}\mathcal{R}}(\mathcal{R}-2\eta
^{-1})+f_{\mathcal{R}\mathcal{T}^{2}}(3p^{2}+\rho^{2}
-\mathcal{T}^{2}))-\gamma^{5}(f_{\mathcal{R}\mathcal{T}
^{2}}(\mathcal{R}
\\\label{44}
&&-2\eta^{-1})+f_{\mathcal{T}^{2}\mathcal{T}^{2}}(3p^{2}
+\rho^{2}-\mathcal{T}^{2}) )\bigg]-\Psi_{,r}=0.
\end{eqnarray}


\begin{thebibliography}{55}

\bibitem{1} Felice, A.D. and Tsujikawa, S.R.: Living Rev. Relativ.
\textbf{13}(2010)3; Nojiri, S. and Odintsov, S.D.: Phys. Rep.
\textbf{505}(2011)59; Bamba, et al.: Astrophys. Space Sci.
\textbf{342}(2012)155.

\bibitem{2} Katirci, N. and Kavuk, M.: Eur. Phys. J. Plus \textbf{129}(2014)163.

\bibitem{2a} Bhattacharjee, S. and Sahoo, P.K.: Eur. Phys. J. Plus \textbf{135}(2020)86.

\bibitem{2b} Singh, K.N. et al.: Phys. Dark Universe \textbf{31}(2021)100774.

\bibitem{2c} Nazari, E.: Phys. Rev. D \textbf{105}(2022)104026.

\bibitem{3a} Roshan, M. and Shojai, F.: Phys. Rev. D \textbf{94}(2016)044002.

\bibitem{3b} Board, C.V. and Barrow, J.D.: Phys. Rev. D \textbf{96}(2017)123517.

\bibitem{3c} Akarsu, O., Katirci, N. and Kumar, S.: Phys. Rev. D \textbf{97}(2018)024011.

\bibitem{3d} Akarsu, O. et al.:  Phys. Rev. D \textbf{98}(2018)063522.

\bibitem{3e} Ranjit, C., Rudra, P. and Kundu, S.: Ann. Phys. \textbf{428}(2021)168432.

\bibitem{3f} Sharif, M. and Naz, S.: Int. J. Mod. Phys. D \textbf{31}(2022)2240008.

\bibitem{4a} Chen, C.Y. and Chen, P.: Phys. Rev. D \textbf{101}(2020)064021.

\bibitem{4b} Akarsu, O., Barrow, J.D. and Uzun, N.M.: Phys. Rev. D \textbf{102}(2020)124059.

\bibitem{4c} Kazemi, A. et al.: Eur. Phys. J. C \textbf{80}(2020)150.

\bibitem{4e} Rudra, P. and Pourhassan, B.: arXiv-2008(2020).

\bibitem{4f} Nazari, E., Sarvi, F. and Roshan, M.: Phys. Rev. D \textbf{102}(2020)064016.

\bibitem{5} Sharif, M. and Gul, M.Z.: Phys. Scr. \textbf{96}(2021)105001.

\bibitem{5a} Sharif, M. and Gul, M.Z.: Pramana J. Phys. \textbf{96}(2022)153.

\bibitem{6} Sharif, M. and Gul, M.Z.: Int. J. Mod. Phys. A \textbf{36}(2021)2150004.

\bibitem{6a} Sharif, M. and Gul, M.Z.: Universe \textbf{7}(2021)154.

\bibitem{6b} Sharif, M. and Gul, M.Z.: Int. J. Geom. Methods Mod. Phys. \textbf{19}(2022)2250012.

\bibitem{6c} Sharif, M., Gul, M.Z.: Chin. J. Phys. \textbf{71}(2021)365.

\bibitem{6d} Sharif, M. and Gul, M.Z.:  Mod. Phys. Lett. A \textbf{37}(2022)2250005.

\bibitem{4g} Yousaf, Z., Bhatti, M.Z. and Farwa, U.: Eur. Phys. J. Plus \textbf{137}(2022)22.

\bibitem{4h} Khodadi, M. and Firouzjaee, J.T.: Phys. Dark Universe \textbf{37}(2022)101084.

\bibitem{10} Flamm, L.: Phys. Z. \textbf{17}(1916)448.

\bibitem{11} Einstein, A. and Rosen, N.: Phys. Rev. \textbf{48}(1935)73.

\bibitem{12} Wheeler, J.A.: Phys. Rev. \textbf{97}(1955)511.

\bibitem{13} Fuller, R.W. and Wheeler, J.A.: Phys. Rev. \textbf{128}(1962)919.

\bibitem{14} Morris, M.S. and Thorne, K.S.: Am. J. Phys. \textbf{56}(1988)395.

\bibitem{15} Kashargin, P.E. and Sushkov, S.V.: Gravit. Cosmol. \textbf{14}(2008)80;
Eiroa, E.F. and Simeone, C.: Phys. Rev. D \textbf{82}(2010)084039.

\bibitem{16} Dzhunushaliev, V. et al.: Phys. Rev. D \textbf{82}(2010)045032.

\bibitem{16a} de Oliveira, P.H.F. et al.: Symmetry \textbf{15}(2023)383.

\bibitem{17} Noether, E.: Tramp. Th. Stat, Phys. \textbf{1}(1918)189.

\bibitem{18} Jamil, M. Mahomed, F.M. and Momeni, D.: Phys.
Lett. B \textbf{702}(2011)315; Basilakos, S., Tsamparlis, M. and
Paliathanasis, A.: Phys. Rev. D \textbf{83}(2011)103512; ibid.
\textbf{84}(2011)123514.

\bibitem{19} Capozziello, S., De Laurentis, M. and Odintsov, S.D.:
Eur. Phys. J. C \textbf{72}(2012)1434; Hussain, I. Jamil, M. and
Mahomed, F.M.: Astrophys. Space Sci. \textbf{337}(2012)373.

\bibitem{20} Motavali, H. and Golshani, M.: Int. J. Mod. Phys. D \textbf{17}(2002)375.

\bibitem{21} Vakili, B.: Phys. Lett. B \textbf{16}(2008)664.

\bibitem{22} Capozziello, S. et al.: Phys. Rev. D \textbf{80}(2009)104030.

\bibitem{23} Capozziello, S., Stabile, A. and Troisi, A.: Class. Quantum Grav. \textbf{25}(2008)085004;
ibid. \textbf{27}(2010)165008.

\bibitem{24} Shamir, M.F., Jhangeer, A. and Bhatti, A.A.: Chin. Phys. Lett.
\textbf{29}(2012)080402.

\bibitem{25} Jamil, M. et al.: Eur. Phys. J. C \textbf{72}(2012)1998.

\bibitem{26} Momeni, D., Myrzakulov, R. and Gudekli, E.:
Int. J. Geom. Methods Mod. Phys. \textbf{12}(2015)1550101.

\bibitem{27} Shamir, M.F. and Ahmad, M.: Eur. Phys. J. C
\textbf{77}(2017)55; Mod. Phys. Lett. A \textbf{32}(2017)1750086.

\bibitem{28} Sharif, M. and Gul, M.Z.: Phys. Scr.
\textbf{96}(2021)025002; Phys. Scr. \textbf{96}(2021)125007; Eur.
Phys. J. Plus \textbf{136}(2021)503; Adv. Astron.
\textbf{2021}(2021)6663502; Chin. J. Phys. \textbf{80}(2022)58.

\bibitem{29} Lobo, F.S.N. et al.: Phys. Rev. D \textbf{80}(2009)104012.

\bibitem{31} Mazharimousavi, S.H. and Halilsoy, M.: Mod. Phys. Lett. A \textbf{31}(2016)1650203.

\bibitem{32} Bahamonde, S. et al.: Phys. Rev. D \textbf{94}(2016)084042.

\bibitem{33} Zubair, M., Waheed, S. and  Ahmed, Y.: Eur. Phys. J. C \textbf{76}(2016)444.

\bibitem{34} Sharif, M. and Nawazish, I.: Ann. Phys \textbf{389}(2018)283.

\bibitem{34a} Sharif, M., Shah, S.A.A. and Bamba, K.: Symmetry \textbf{10}(2018)153;
Sharif, M. and Saba, S.: Symmetry \textbf{11}(2019)92.

\bibitem{35} Mustafa, G. et al.: Int. J. Geom. Methods Mod. Phys. \textbf{17}(2020)2050103.

\bibitem{36} Shamir, M.F. and Fayyaz, I.: Eur. Phys. J. C \textbf{80}(2020)1102.

\bibitem{36a} Hassan, Z., Mustafa, G. and Sahoo, P.K.: Symmetry \textbf{13}(2021)1260.

\bibitem{37} Malik, A. et al.: Chin. Phys. C \textbf{46}(2022)095104.

\bibitem{38} Ellis, G.F.R., Maartens, R. and MacCallum, M.A.H., \emph{Relativistic Cosmology}
(Cambridge University Press, Cambridge, 2012).

\bibitem{39} Roshan, M. and Shojai, M.: Phys. Rev. D \textbf{94}(2016)044002.

\bibitem{40} Cataldo, M., Liempi, L. and Rodriguez, P.: Phys. Lett. B \textbf{757 }(2016)130.

\bibitem{41} Gyulchev, G. et al.: Eur. Phys. J. C \textbf{78}(2018)544.
\bibitem{41a} Abreu, H. et al.: Class. Quantum Gravity \textbf{24}(2007)4631.

\bibitem{42} Heintzmann, H. and Hillebrandt, W.: Astron. Astrophys.
\textbf{38}(1975)51.

\bibitem{43} Capozziello, S., De Laurentis, M.: Phys. Rep. \textbf{509}(2011)167.
Shamir, M.F. and Ahmad, M.: Eur. Phys. J. C \textbf{77}(2017)674;
Deb, D. et al.: Ann. Phys. \textbf{387}(2017)239; Sharif, M.,
Siddiqa, A.: Int. J. Mod. Phys. D \textbf{27}(2018)1850065.

\bibitem{44} Fayyaz, I. and Shamir, M.F.: Chin. J. Phys. \textbf{66}(2020)553.

\bibitem{45} Shamir, M.F. and Fayyaz, I.: Eur. Phys. J. C \textbf{80}(2020)1102.

\end{thebibliography}
\end{document}